\documentclass[12pt]{article}
\usepackage{epsfig,a4}

 \advance\oddsidemargin by -1in
 \advance\evensidemargin
 by -1in
 \marginparsep
 0.4cm
 \advance\topmargin by
 -0.0in
 \makeindex

 \newcommand\la{\langle}
 \newcommand\ra{\rangle}
 \newcommand\beq{\begin{equation}}
 
 \newcommand\eeq{\end{equation}}                                               
 \newcommand\beqn{\begin{eqnarray}}
 \newcommand\eeqn{\end{eqnarray}}
 \newcommand{\doublespace} {
 \renewcommand{\baselinestretch} {1.6}
 \large\normalsize} 

\def\doublespace{\def\baselinestretch{1.6}\large\normalsize}

\def\BA{\begin{eqnarray}}
\def\BE{\begin{equation}}
\def\BF{\begin{figure}[htb]}
\def\BT{\begin{table}[htb]}
\def\EA{\end{eqnarray}}
\def\EE{\end{equation}}
\def\EF{\end{figure}}
\def\ET{\end{table}}

\def\la{\langle}
\def\ra{\rangle}
\def\mb{\,\mbox{mb}}
\def\fm{\,\mbox{fm}}
\def\GeV{\,\mbox{GeV}}

\def\Pom{{\bf I\!P}}

\def\lsim{\mathrel{\rlap{\lower4pt\hbox{\hskip1pt$\sim$}}
    \raise1pt\hbox{$<$}}}         
\def\gsim{\mathrel{\rlap{\lower4pt\hbox{\hskip1pt$\sim$}}
    \raise1pt\hbox{$>$}}}         

\begin{document} 
\vspace*{3cm}
 
\date{today} 

\begin{center}
{\Large\bf Color Transparency versus Quantum Coherence

\medskip
in Electroproduction of Vector Mesons off Nuclei}

\end{center}

 
\begin{center}
 
\vspace{0.5cm}
 {\large B.Z.~Kopeliovich$^{1,2,3}$, J.~Nemchik$^{4}$, 
A.~Sch\"afer$^{2}$ and A.V.~Tarasov$^{1,2,3}$}
 \\[1cm]
 $^{1}${\sl Max-Planck Institut f\"ur Kernphysik, Postfach 103980, 69029
Heidelberg, Germany}\\[0.2cm]
 $^{2}${\sl Institut f\"ur Theoretische Physik der Universit\"at, 93040
Regensburg, Germany} \\[0.2cm]
 $^{3}${\sl Joint Institute for Nuclear Research, Dubna, 141980 Moscow
Region, Russia}\\[0.2cm]
 $^{4}${\sl Institute of Experimental Physics SAV, Watsonova 47, 
04353 Kosice, Slovakia}
 
\end{center}

\vspace{1cm}
 
\begin{abstract}

So far no theoretical tool for the comprehensive description of exclusive
electroproduction of vector mesons off nuclei at medium energies has been
developed. We suggest a light-cone QCD formalism which is valid at any
energy and incorporates formation effects (color transparency), the
coherence length and the gluon shadowing. At medium energies color
transparency (CT) and the onset of coherence length (CL) effects are not
easily separated. Indeed, although nuclear transparency measured by the
HERMES experiment rises with $Q^2$, it agrees with predictions of the
vector dominance model (VDM) without any CT effects. Our new results and
observations are: (i) the good agreement with the VDM found earlier is
accidental and related to the specific correlation between $Q^2$ and CL
for HERMES kinematics; (ii) CT effects are much larger than have been
estimated earlier within the two channel approximation. They are even
stronger at low than at high energies and can be easily identified by
HERMES or at JLab;  (iii) gluon shadowing which is important at high
energies is calculated and included; (iv) our parameter-free calculations
explain well available data for variation of nuclear transparency with
virtuality and energy of the photon; (v) predictions for
electroproduction of $\rho$ and $\phi$ are provided for future
measurements at HERMES and JLab.

\end{abstract}
 
\doublespace
 
\newpage
 
\section{Introduction: interplay of absorption and shadowing} 
\label{intro}

\subsection{Color transparency}\label{ct}

The nuclear medium is more transparent for colorless hadronic wave
packets than predicted by the Glauber model. One can treat this
phenomenon either in the hadronic basis as a results of Gribov's
inelastic corrections \cite{gribov}, or in QCD as a result of color
screening \cite{zkl,bbgg}, an effect called color transparency (CT) (see
also the review \cite{jpr}). Although the two approaches are
complementary, the latter interpretation is more intuitive and
straightforward. Indeed, a point-like colorless object cannot interact
with external color fields, therefore its cross section vanishes
$\sigma(r)\propto r^2$ at $r \to 0$ \cite{zkl}. This fact naturally
explains the correlation between the cross sections of hadrons and their
sizes \cite{gs,hp,p}. When a colorless wave packet propagates through a
nucleus, the fluctuations with small size have an enhanced survival
probability which leads to a non-exponential attenuation $\propto 1/L$
\cite{zkl}, where $L$ is the path length in nuclear matter.

Diffractive electroproduction of vector mesons off nuclei is affected by
shadowing and absorption which are different phenomena. Final state
absorption of the produced meson exists even in the classical
probabilistic approach which relates nuclear suppression to the
survival probability $W(z,b)$ of the vector meson produced at the
point with longitudinal coordinate $z$ and impact parameter $\vec b$,
 \beq
W(z,b) = exp\left[-\,\sigma_{in}^{VN}\,
\int\limits_{z}^{\infty}
dz^\prime\,\rho_A(b,z^\prime)\right]\ ,
\label{10}
\eeq
 where $\rho_A(b,z)$ is the nuclear density and $\sigma_{in}^{VN}$ is
the inelastic $VN$ cross section.

Going beyond the VDM one realizes that the diffractive process initiating 
the
production of the vector meson on a bound nucleon is $\gamma^*\,N \to
\bar qq\,N$ (with possible glue). A photon of high virtuality $Q^2$ is
expected to produce a pair with a small $\sim 1/Q^2$ transverse
separation\footnote{In fact, the situation is somewhat more complicated.
For very asymmetric pairs when the $q$ or $\bar q$ carry almost the whole
photon momentum, the pair can have a large separation, see
Sect.~\ref{gamma-wf}}. The basic idea of CT is that such 
small size
should lead to a vanishing absorption when the colorless $\bar qq$ wave
packet propagates through the nucleus. However, the pair may evolve in
size during the time of propagation due to transverse motion of the
quarks. Besides, the medium filters out large size configurations which
have larger absorption cross section\footnote{Absorption does not mean
disappearance or stopping of the quarks. High energy partons usually lose
only a very small (energy independent)  fraction of their energy,
primarily via soft QCD processes. Absorption means color-exchange
interaction which switches from the exclusive channel to an inclusive
process.}, an effect known as color filtering. Eventually, the resulting
distribution amplitude of the $\bar qq$ wave packet must be projected
onto the wave function of the $V$ meson.

The time scale characterizing the evolution of the $\bar qq$ wave packet
can be estimated based on the uncertainty principle. One cannot decide
whether the ground state $V$ is produced or the next excited state
$V^\prime$, unless the process lasts longer than the inverse mass
difference. In the rest frame of the nucleus this formation time is
Lorentz dilated,
 \beq
t_f = \frac{2\,\nu}
{\left.m_V^\prime\right.^2 - m_V^2}\ ,
\label{20}
 \eeq
 where $\nu$ is the photon energy.

A rigorous quantum-mechanical description of the pair evolution was
suggested in \cite{kz-91} and is based on the light-cone Green function
technique. This approach is presented below in Sect.~\ref{lc}.

A complementary description of the same process in the hadronic basis
looks quite different \cite{hk-97}. The incident photon may produce
different states on a bound nucleon, the $V$ meson ground state or an
excited state. Those states propagate through the nucleus experiencing
multiple diagonal and off-diagonal diffractive interactions, and
eventually the ground state is detected. According to quark-hadron
duality we expect these two descriptions to be equivalent. In practice,
however, neither of them can be calculated exactly, and therefore each
has advantages and shortcomings.  For example, electroproduction of light
vector mesons on a nucleon cannot be calculated perturbatively without
reservations, while in the hadronic basis one can make use of
experimental data which include all nonperturbative effects.  On the
other hand, for excited meson states no data are available for the
diagonal and off-diagonal diffractive amplitudes which one can estimate
in the quark representation. The two approaches are complementary, they
rely on different approximations and their comparison may provide a scale
for the theoretical uncertainty involved.

\subsection{Effects of coherence: shadowing of quarks and
gluons}\label{shad}

Another phenomenon, shadowing, is also known to cause nuclear
suppression. In contrast to final state absorption, it is a pure
quantum-mechanical effect which results from destructive interference of
the amplitudes for which the interaction takes place on different bound
nucleons. It can be interpreted as a competition between the different
nucleons participating in the reaction: since the total probability
cannot exceed one, each participating nucleon diminishes the chances of
others to contribute to the process.

The cross section of photoproduction is very small since it includes the
fine structure constant. Applying the Glauber formula one should expect
no visible shadowing. However, this is true only at low energies. It has
been realized back in the 60s (see the review \cite{bauer}) that the
photon interacts via its hadronic fluctuations. Therefore, if a
fluctuation can propagate over a distance comparable or longer than the
nuclear radius, it may interact with a large hadronic cross section which
causes shadowing. The small probability to create such a fluctuation
enters only once, otherwise the fluctuation interacts strongly. Thus, the
fluctuation lifetime provides the time scale which controls shadowing.
Again, it can be estimated relying on the uncertainty principle and
Lorentz time dilation as,
 \beq
t_c = \frac{2\,\nu}{Q^2 + m_V^2}\ .
\label{30}
 \eeq 
 It is usually called coherence time, but we also will use the term
coherence length (CL), since light-cone kinematics is assumed, $l_c=t_c$
(similarly, for formation length $l_f=t_f$). CL is related to the
longitudinal momentum transfer $q_c=1/l_c$ in $\gamma^*\,N \to V\,N$,
which controls the interference of the production amplitudes from
different nucleons.

Initial state shadowing indeed has been observed in many reactions where
no final state absorption is expected, for example in the total
photoabsorption cross section on nuclei (see \cite{bauer}), the inclusive
deep-inelastic cross section \cite{nmc,e665}, the total neutrino-nucleus
cross section \cite{k-neutrino}, the Drell-Yan reaction of dilepton
production \cite{e772,e866}, etc. In the case of electroproduction of
vector mesons off nuclei shadowing and absorption happen with the same
cross section which makes it difficult to
disentangle the two sources of nuclear suppression. Nevertheless, it is
easy to identify the difference in the two limiting cases \cite{kz-91}
which we illustrate for the example of the vector dominance model (VDM).
The first case is the limit of small $l_c$, shorter than the mean
internucleon spacing $\sim 2\,fm$. In this case only final state
absorption matters. The ratio of the quasielastic (or incoherent)
$\gamma^*\, A \to V\,X$ and $\gamma^*\, N \to V\,X$ cross sections,
usually called nuclear transparency, reads,
 \beqn
Tr_A^{inc}\Bigr|_{l_c\ll R_A} &\equiv& 
\frac{\sigma_V^{\gamma^*A}}
{A\,\sigma_V^{\gamma^*N}} =
\frac{1}{A}
\,\int d^2b\,
\int\limits_{-\infty}^{\infty}
dz\,\rho_A(b,z)\,
\exp\left[-\sigma^{VN}_{in}
\int\limits_z^{\infty} dz'\,
\rho_A(b,z')\right]\nonumber\\
&=& \frac{1}{A\,\sigma^{VN}_{in}}\,
\int d^2b\,\left\{1 - 
\exp\Bigl[-\sigma^{VN}_{in}\,T(b)\Bigr]\right\}=
\frac{\sigma^{VA}_{in}}{A\,\sigma^{VN}_{in}}\ .
\label{40}
 \eeqn

In the limit of long $l_c$ it takes a different form,
 \beq
Tr_A^{inc}\Bigr|_{l_c\gg R_A} = 
\int d^2b\,T_A(b)\,
\exp\left[-\sigma^{VN}_{in}\,T_A(b)\right]\ ,
\label{50}
 \eeq where we assume $\sigma^{VN}_{el} \ll \sigma^{VN}_{in}$ for the
sake of simplicity.  $T_A(b)$ is the nuclear thickness function
 \beq
T_A(b) = \int\limits_{-\infty}^{\infty} dz\,\rho_A(b,z)\ .
\label{60}
 \eeq 
 The exact expression beyond VDM which interpolates between the two
regimes (\ref{40}) and (\ref{50}) can be found in \cite{hkn}.

One can see that the $V$ meson attenuates along the whole nucleus
thickness in Eq.~(\ref{50}), but only along roughly half of that length
in (\ref{40}).  This confirms our conjecture that nuclear shadowing also
contributes to (\ref{50}) increasing suppression. This may be also
interpreted as an analog of the quark nuclear shadowing measured in DIS
off nuclei, but the absorption effects make this analogy rather shaky.

Gluon shadowing also suppresses electroproduction of $V$ mesons.  
Different (but equivalent) descriptions of gluon shadowing are known. In
the infinite momentum frame of the nucleus it looks like fusion of gluons
which overlap in longitudinal direction at small $x$, leading to a
reduction of gluon density. In the rest frame of the nucleus the same
phenomenon looks as a specific part of Gribov's inelastic corrections
\cite{gribov}. The lowest order inelastic correction related to
diffractive dissociation $V\,N \to X\,N$ \cite{kk} contains PPR and PPP
contributions (in terms of the triple-Regge phenomenology, see
\cite{kklp}). The former is related to the quark shadowing already
discussed above, while the latter, the triple-Pomeron term, corresponds
to gluon shadowing. Indeed, only diffractive gluon radiation can provide
the $M_X$ dependence $d\sigma_{dd}/dM^2_X \propto 1/M_X^2$ of the
diffractive dissociation cross section.

In terms of the light-cone QCD approach the same process is related to
the inclusion of higher Fock components, $|\bar qq\,nG\ra$, containing
gluons \cite{al}. Such fluctuations might be quite heavy compared to the
simplest $|\bar qq\ra$ fluctuation, therefore, they have a shorter
lifetime \cite{krt2} and need a higher energy to be relevant.

\subsection{Outline of the paper}

In Sect.~\ref{lc} we present the light-cone (LC) approach to diffractive
electroproduction of vector mesons in the rest frame of the nucleon
target. The central issue of this approach, the universal interaction 
cross section for a colorless quark-antiquark dipole and a 
nucleon, is presented in Sect.~\ref{dipole-cross}. It cannot be 
reliably evaluated theoretically and is fitted to the data for the proton 
structure function in a wide range of $x_{Bj}$ and $Q^2$.

The LC wave function for a quark-antiquark fluctuation of the virtual
photon is presented in Sect.~\ref{gamma-wf} for both, free and
interacting $\bar qq$ pairs. In the latter case we apply the LC Green
function approach and introduce into the two-dimensional Schr\"odinger
equation a nonperturbative real LC potential describing the $\bar qq$
interaction. The model for the LC wave of a vector meson is described in
Sect.~\ref{rho-wf}.

As a rigorous test of the model we calculate in Sect.~\ref{data-N} the
cross section of elastic electroproduction of $\rho$ and $\phi$ mesons
off a nucleon target.  These parameter-free calculations reproduce both
energy and $Q^2$ dependence remarkably well, including the absolute
normalization. Since we use the nonperturbative LC photon wave function
it is legitimate to do calculations down to $Q^2=0$. Agreement with data
for real photoproduction of $\rho$ and $\phi$ is also good.

Sect.~\ref{q-qbar} is devoted to incoherent production of vector mesons
off nuclei. In Sect.~\ref{green-f} the Green function describing
propagation of a $\bar qq$ in the nuclear medium is modified to
incorporate absorption. This is done by introducing an imaginary part of
the potential into the two-dimensional LC Schr\"odinger equation for the
Green function. Different limiting cases of short and long coherence and
formation lengths are considered. The central results of the paper is
Eq.~(\ref{520}) for the cross section of incoherent vector meson
production in the most general case.  Numerical calculations and
comparison with available data are presented in Sect.~\ref{incoh-data}.
Nuclear transparency turns out to be a result of a complicated interplay
between coherence and formation length effects. Although variation of
$l_c$ with $Q^2$ can mimic CT at medium and low energies, one can map
experimental events in $Q^2$ and $\nu$ in such a way as to keep
$l_c=const$. Unexpectedly, the exact solution found in the present work
is very different from the two-coupled-channel approximation of
\cite{hk-97} and predicts a much more pronounced effect of CT. This makes
it feasible to find a clear signal of CT effects in exclusive production
of $\rho$ mesons in the current and planned experiments at HERMES and
JLab.

Coherent production of vector mesons off nuclei leaving the nucleus
intact is studied in Sect.~\ref{coh}. The formalism described in
Sect.~\ref{coh-form} is simpler than in the case of incoherent
production. The detailed calculations and the comparison with data are
presented in Sect.~\ref{coh-data}. The effect of CT on the $Q^2$
dependence of nuclear transparency at $l_c=const$ is weaker than in the
case of incoherent production and is difficult to be detected at low
energies since the cross section is small. Our results for the
differential cross section of coherent production of $\rho$ demonstrate
also a weak sensitivity to the CT effects.

Besides CL, there are other effects considered in Sect.~\ref{pitfalls}
which can mimic the phenomenon of CT. First, the standard lowest order
inelastic corrections well fixed by available data are known to make the 
nuclear medium more transparent at higher energies. Since $\nu$ is a
rising function of $Q^2$ at fixed $l_c$, nuclear transparency increases
with $Q^2$.  These corrections are estimated in Sect.~\ref{standard-corr}
and the effect is found to be too weak to mock CT. Another source of
rising $Q^2$ dependence of the nuclear transparency is the finite $\rho$
lifetime which might be important at low energies. This effect evaluated
in Sect.~\ref{lifetime} is also found to be negligibly small.

Exclusive production of vector mesons at high energies is controlled by
the small-$x_{Bj}$ physics, and gluon shadowing becomes an important
phenomenon. It affects the cross section of incoherent vector meson
production in a two-fold way. While the production of $V$ on a bound
nucleon is suppressed, the nuclear medium becomes more transparent
enhancing the survival probability of $\bar qq$ wave packets traveling
through the nucleus.  At the same time, the cross section of coherent
production can be only diminished.  In Sect.~\ref{glue-shadow} gluon
shadowing is calculated and included in the calculations for nuclear
transparency.

The results of the paper are summarized and discussed in
Sect.~\ref{conclusions}. An optimistic prognosis for the CT discovery
potential of future experiments at HERMES and JLab is made.

\section{Light-cone dipole phenomenology for elastic photoproduction 
of vector mesons \boldmath$\gamma^{*}N\to V~N$}\label{lc}

In the light-cone dipole approach the amplitude of a diffractive process
is treated as elastic scattering of a $\bar qq$ fluctuation of the
incident particle.  The elastic amplitude given by convolution of the
universal flavor independent dipole cross section for the $\bar qq$
interaction with a nucleon, $\sigma_{\bar qq}$, which is introduced in
\cite{zkl}, and the initial and final wave functions \cite{zkl}.  Thus,
the forward production amplitude for the exclusive photo- or
electroproduction of vector mesons $\gamma^{*}N\to V~N$ can be
represented in the form
 \BE
{\cal M}_{\gamma^{*}N\rightarrow VN}(s,Q^{2}) =
\langle V |\sigma_{\bar qq}^N({\vec{\rho}},s)|\gamma^{*}\rangle=
\int\limits_{0}^{1} d\alpha \int d^{2}{{r}}\,\,
\Psi_{V}^{*}({\vec{r}},\alpha)\,   
\sigma_{\bar qq}({\vec{r}},s)\,  
\Psi_{\gamma^{*}}({\vec{r}},\alpha,Q^2)\,
\label{120}
 \EE
 with the normalization 
 \beq
\left.\frac{d\sigma}{dt}\right|_{t=0} =
\frac{|{\cal M}|^{2}}{16\,\pi}.
\label{125}
 \eeq

In order to calculate the photoproduction amplitude one needs to know the
following ingredients of Eq.~(\ref{120}): (i)  the dipole cross section
$\sigma_{\bar qq}({\vec{r}},s)$ which depends on the $\bar qq$
transverse separation $\vec{r}$ and the c.m. energy squared $s$. (ii)  
The light-cone (LC)  wave function of the photon
$\Psi_{\gamma^{*}}({\vec{r}},\alpha,Q^2)$ which also depends on the
photon virtuality $Q^2$ and the relative share $\alpha$ of the photon
momentum carried by the quark. (iii) The LC wave function
$\Psi_{V}({\vec{r}},\alpha)$ of the vector meson. They are presented in
the following sections.

Note that in the LC formalism the photon and meson wave functions contain
also higher Fock states $|\bar qq\ra$, $|\bar qqG\ra$, $|\bar qq2G\ra$,
etc. Should one add their contribution to Eq.~(\ref{125})? A word of
caution is in order.  The energy dependence of the total cross section
including the dipole one as is given in Eq.~(\ref{120}), originates from
inelastic collisions with gluon bremsstrahlung, a process related to the
forward elastic amplitude via unitarity.  Those inelastic collisions also
can be described in terms of the Fock components containing gluons.  
Thus, one would double count if both the energy dependent dipole cross
section and the higher Fock states were included.  Either one should rely
upon the Fock state decomposition treating interaction of each of them as
energy independent, or one should restrict ones consideration to the
lowest $|\bar qq\ra$ component, but implicitly incorporate the effects of
higher Fock states into the energy dependence of the dipole cross section
$\sigma_{\bar qq}$. We stand with the latter approach in the present
paper.

However, as for nuclear targets, one must explicitly include into ones
consideration the higher Fock states because their eikonalization leads
to gluon shadowing. We come back to this problem in
Sect.~\ref{glue-shadow}.

\subsection{Phenomenological dipole cross section}\label{dipole-cross}

The cross section $\sigma_{\bar qq}(\vec r,s)$ for the interaction of a
$\bar qq$ dipole of transverse separation $\vec r$ with a nucleon, first
introduced in \cite{zkl}, is a flavor independent universal function of
$\vec{r}$ and energy. It allows to describe in a uniform way various
high energy processes.  This cross section still cannot be predicted
reliably because of poorly known higher order pQCD corrections and
nonperturbative effects. However, it is known to vanish quadratically
$\sigma_{\bar qq}(r,s)\propto r^2$ as $r\rightarrow 0$ due to color
screening, a property usually called color transparency. On the other
hand, one may expect the dipole cross section to level off at large
separations. This may happen if the quark density in the proton already
saturates in the $x_{Bj}$ range of HERA \cite{levin,gbw}. Whether this
was already observed at HERA in the small $x_{Bj}$ domain is difficult to
say. One can fit the data perfectly either assuming saturation
\cite{gbw}, or with the pure DGLAP evolution. One can interpret the
leveling off of the dipole cross section not only in terms of saturated
parton density (these two are {\it not} identical).  Another scenario
relates the flat behavior of $\sigma_{\bar qq}(r,s)$ at large $r$ to the
averaged gluon propagation length $r_0$. For $r^2\gg r_0^2$ one arrives
in this case at the additive quark model: the dipole cross section is a
sum of quark-quark cross sections, i.e. the $\sigma_{\bar qq}(r)$ levels
off at large separations.

At small separations the dipole cross section should be a function of $r$
and $x_{Bj}\sim 1/(r^2s)$ to reproduce Bjorken scaling. A corresponding
simple and popular parameterization has been suggested in \cite{gbw}. It
well describes data for DIS at small $x$ and medium and high $Q^2$.
However, at small $Q^2$ it cannot be correct since it predicts energy
independent hadronic cross sections. Besides, $x_{Bj}$ is not any more a
proper variable at small $Q^2$ and should be replaced by energy.  Since
we want our approach to be valid down to the limit of real
photoproduction, we choose the parametrization suggested in \cite{kst2}
which is similar to one in \cite{gbw}, but contains an explicit
dependence on energy,
 \BA
  \sigma_{\bar qq}(r,s) &=& \sigma_0(s) 
\left[1 - e^{- r^2/r_{0}^2(s)}
  \right]\ .
  \label{130}
 \EA
 It correctly reproduces the hadronic cross sections for the choice
 \BE
  \sigma_0(s) = \sigma^{\pi p}_{tot}(s)
  \left[1+\frac38 \frac{r_{0}^2(s)}
{\left<r^2_{ch}\right>}\right]\mb\ ,\\
  \label{140}
 \EE
 \BE
  r_0(s)   = 0.88 \left(\frac{s}{s_0}\right)^{\!\!-0.14}  \fm\ .
  \label{150}
 \EE
 Here $\left<r^2_{ch}\right>=0.44\fm^2$ is the mean pion charge radius
squared; $s_0 = 1000\GeV^2$.  The cross section $\sigma^{\pi p}_{tot}(s)$
was fitted to data in \cite{dl,pdt},
 \BE
\sigma^{\pi p}_{tot}(s) = 
23.6\,\left(\frac{s}{s_0}\right)^{\!\!0.079} 
+ 0.032\ \left(\frac{s}{s_0}\right)^{-0.45}\,\mb\ .
\label{145}
 \EE
 It represents the Pomeron and Reggeon parts corresponding to exchange in
of gluons and $\bar qq$, respectively.  Only the former has been used in
the dipole cross section Eq.~(\ref{130}) to fit the data for the proton
structure function at small $x_{Bj}$.  Unfortunately, the Reggeon part of
the dipole cross section is poorly known. To the best of our knowledge no
phenomenology of it has been developed so far. The energy dependence of
the Reggeon dipole cross section at small $Q^2$ (or $x$-dependence at
high $Q^2$) dictated by Regge phenomenology is approximately $\propto
1/\sqrt{s}$ ($\propto \sqrt{x}$). Thus, we can expect $Q^2$ independence
of the the exponent in the second term on the r.h.s of Eq.~(\ref{145}).  
The $r$ dependence of this term in the dipole cross section is less
known. In order to reproduce Bjorken scaling we must assume that the
Regge term vanishes at $r\to 0$ in the same way as the Pomeron part of
the cross section, $\sigma_R(r,s)\propto r^2$. For the sake of simplicity
we therefore assume the same parameterization as is used for the Pomeron
part, Eq.~(\ref{130}). Then, one can just incorporate the Reggeon term
into $\sigma_0(s)$ as is done in Eq.~(\ref{145}).

Incorporating Reggeons into the LC dipole formalism for nuclear shadowing
one should be careful with the treatment of multiple interaction terms
which have a nonplanar nature \cite{gribov-nonplanar}. The Reggeon
exchange described by planar graphs should not participate in the
multiple-scattering expansion of the eikonal exponential.

The dipole cross section Eqs.(\ref{130}) -- (\ref{145}) provides the
imaginary part of the elastic amplitude.  It is known, however, that the
energy dependence of the total cross section generates also a real part
\cite{bronzan},
 \beq
\sigma_{\bar qq}(r,s) \Rightarrow
\left(1-i\,\frac{\pi}{2}\,\frac{\partial}
{\partial\ln(s)}\right)\,
\sigma_{\bar qq}(r,s)
\label{real-part}
 \eeq
 The energy dependence of the dipole cross section Eq.~(\ref{130}) is
rather steep at small $r$ leading to a large real part which should not
be neglected. For instance, the photoproduction amplitude of $\gamma N\to
J/\Psi N$ rises $\propto s^{0.2}$ and the real-to-imaginary part ratio is
over $30\%$. At medium energies also the Reggeon contribution to
electroproduction of light mesons contributes to the real part of the
elastic amplitude. The replacement Eq.~(\ref{real-part}) takes care of it
as well, and we use this form everywhere in what follows, unless
specified otherwise.

Note that the improvement compared to \cite{gbw} at large separations
leads to a worse description of the short-distance part of the dipole
cross section which is responsible for the behavior of the proton
structure function at large $Q^2$. To satisfy Bjorken scaling the dipole
cross section at small $r$ must be a function of the product $s r$ which
is not the case for the parametrization in Eq.~(\ref{130}). Indeed, the
form of Eq.~(\ref{130}) successfully describes data for DIS at small $x$
only up to $Q^2\approx 10\GeV^2$, and does a poor job at larger values of
$Q^2$. Nevertheless, this interval of $Q^2$ is sufficient for the purpose
of the present paper which is focused on production of light vector
mesons at small and moderate $Q^{2}\lsim 10$\,\GeV$^{2}$.

\subsection{The \boldmath$\bar qq$ wave function of the 
photon}\label{gamma-wf}

The perturbative distribution amplitude (``wave function'') of the $\bar
qq$ Fock component of the photon is well known \cite{lc,bks-71,nz-91},
and for transversely (T) and longitudinally (L) polarized photons it has
the form,
 \BE
\Psi_{\bar qq}^{T,L}({\vec{r}},\alpha) =
\frac{\sqrt{N_{C}\,\alpha_{em}}}{2\,\pi}\,\,
Z_{q}\,\bar{\chi}\,\hat{O}^{T,L}\,\chi\, 
K_{0}(\epsilon\,r)
\label{70}
 \EE
 where $\chi$ and $\bar{\chi}$ are the spinors of the quark and
antiquark, respectively; $Z_{q}$ is the quark charge; $N_{C} = 3$ is the
number of colors. $K_{0}(\epsilon r)$ is a modified Bessel
function with 
 \BE
\epsilon^{2} =
\alpha\,(1-\alpha)\,Q^{2} + m_{q}^{2}\ ,
\label{80}
 \EE
 where $m_{q}$ is the quark mass, and $\alpha$ is the fraction of the
LC momentum of the photon carried by the quark. The operators
$\widehat{O}^{T,L}$ read,
 \BE 
\widehat{O}^{T} = m_{q}\,\,\vec{\sigma}\cdot\vec{e} +
i\,(1-2\alpha)\,(\vec{\sigma}\cdot\vec{n})\,
(\vec{e}\cdot\vec{\nabla}_r) + (\vec{\sigma}\times
\vec{e})\cdot\vec{\nabla}_r\ ,
 \label{90}
 \EE
 \BE
\widehat{O}^{L} =
2\,Q\,\alpha (1 - \alpha)\,(\vec{\sigma}\cdot\vec{n})\ .
\label{100}
 \EE
 Here $\vec\nabla_r$ acts on transverse coordinate $\vec r$;
$\vec{e}$ is the polarization vector of the photon and $\vec{n}$ is a unit
vector parallel to the photon momentum.

The transverse $\bar qq$ separation is controlled by the distribution 
amplitude Eq.~(\ref{70}) with the mean value,
 \BE
\la r\ra \sim \frac{1}{\epsilon} = 
\frac{1}{\sqrt{Q^{2}\,\alpha\,(1-\alpha) + m_{q}^{2}}}\,.
\label{212}
 \EE
 In pQCD the quarks are treated as free, and one may wonder why they do
not fly apart but form a wave packet of finite size. It is interference
of $\bar qq$ waves produced at different points which keeps the
transverse separation finite. To reach a large separation the $\bar qq$
pair must be produced sufficiently long in advance, longer than the
coherence time Eq.~(\ref{30}), such that fluctuations loose coherence.  
Treating the coherence time as lifetime of the fluctuation, one can also
say that the fluctuation does not have enough time to fly apart.

 For very asymmetric $\bar qq$ pairs with $\alpha$ or $(1-\alpha) \lsim
m_q^2/Q^2$ the mean transverse separation $\la r\ra \sim 1/m_q$ becomes
huge since one must use current quark masses within pQCD. A popular
recipe to fix this problem is to introduce an effective quark mass
$m_{eff}\sim \Lambda_{QCD}$ which should represent the nonperturbative
interaction effects between $q$ and $\bar q$. It is more consistent,
however, and straightforward to introduce this interaction explicitly.
The corresponding phenomenology based on the light-cone Green function
approach has been developed in \cite{kst2}.

The Green function $G_{\bar qq}(z_1,\vec r_1;z_2,\vec r_2)$ 
describes the 
propagation of an interacting $\bar qq$ pair between points with
longitudinal coordinates $z_{1}$ and $z_{2}$ and with initial and final
separations $\vec r_1$ and $\vec r_2$. This
Green function satisfies the 
two-dimensional Schr\"odinger equation, 
 \BE
i\frac{d}{dz_2}\,G_{\bar qq}(z_1,\vec r_1;z_2,\vec r_2)=
\left[\frac{\epsilon^{2} - \Delta_{r_{2}}}{2\,\nu\,\alpha\,(1-\alpha)}
+V_{\bar qq}(z_2,\vec r_2,\alpha)\right]
G_{\bar qq}(z_1,\vec r_1;z_2,\vec r_2)\ .
\label{250}  
 \EE
 Here $\nu$ is the photon energy. The Laplacian $\Delta_{r}$ acts on
the coordinate $r$.  

The imaginary part of the LC potential $V_{\bar
qq}(z_2,\vec r_2,\alpha)$ in (\ref{250}) is responsible for
attenuation of the $\bar qq$ in the medium, while the real part
represents the interaction between the $q$ and $\bar{q}$.  
This potential is supposed to provide the correct LC wave functions of 
vector mesons. For the sake of simplicity we use  the oscillator form 
of the potential,
 \BE  
{\rm Re}\,V_{\bar qq}(z_2,\vec r_{2},\alpha) =
\frac{a^4(\alpha)\,\vec r_{2}\,^2} 
{2\,\nu\,\alpha(1-\alpha)}\ ,
\label{260} 
 \EE
 which leads to a Gaussian $r$-dependence of the LC wave function of the
meson ground state.  The shape of the function $a(\alpha)$ will be
discussed in the next section.

 In this case equation (\ref{250}) has an analytical solution, the
harmonic oscillator Green function \cite{fg},
 \BA 
G_{\bar qq}(z_1,\vec r_1;z_2,\vec r_2) =
\frac{a^2(\alpha)}{2\;\pi\;i\;
{\rm sin}(\omega\,\Delta z)}\, {\rm exp}
\left\{\frac{i\,a^2(\alpha)}{{\rm sin}(\omega\,\Delta z)}\,
\Bigl[(r_1^2+r_2^2)\,{\rm cos}(\omega \;\Delta z) -
2\;\vec r_1\cdot\vec r_2\Bigr]\right\}
\nonumber\\ \times {\rm exp}\left[- 
\frac{i\,\epsilon^{2}\,\Delta z}
{2\,\nu\,\alpha\,(1-\alpha)}\right] \ , 
\label{270} 
 \EA
where $\Delta z=z_2-z_1$ and 
 \BE \omega = \frac{a^2(\alpha)}{\nu\;\alpha(1-\alpha)}\ .
\label{280} 
 \EE
 The boundary condition is $G_{\bar
qq}(z_1,\vec r_1;z_2,\vec r_2)|_{z_2=z_1}=
\delta^2(\vec r_1-\vec r_2)$.

The probability amplitude to find the $\bar qq$ fluctuation of a photon
at the point $z_2$ with separation $\vec r$ is given by an integral
over the point $z_1$ where the $\bar qq$ is created by the photon with
initial separation zero,
 \BE
\Psi^{T,L}_{\bar qq}(\vec r,\alpha)=
\frac{i\,Z_q\sqrt{\alpha_{em}}}
{4\pi\,E\,\alpha(1-\alpha)} 
\int\limits_{-\infty}^{z_2}dz_1\,
\Bigl(\bar\chi\;\widehat O^{T,L}\chi\Bigr)\,
G_{\bar qq}(z_1,\vec r_1;z_2,\vec r)
\Bigr|_{r_1=0}\ .
\label{290}
 \EE
 The operators $\widehat O^{T,L}$ are defined in Eqs.~(\ref{90}) and
(\ref{100}). Here they act on the coordinate $\vec r_1$.

If we write the transverse part  as
 \BE
\bar\chi\;\widehat O^{T}\chi= A+\vec B\cdot\vec\nabla_{r_1}\ ,
\label{300}
 \EE
 then the distribution functions read,   
 \BE
\Psi^{T}_{\bar qq}(\vec r,\alpha) =
Z_q\sqrt{\alpha_{em}}\,\left[A\,\Phi_0(\epsilon,r,\lambda)
+ \vec B\,\vec\Phi_1(\epsilon,r,\lambda)\right]\ ,
\label{310}
 \EE
 \BE
\Psi^{L}_{\bar qq}(\vec r,\alpha) =
2\,Z_q\sqrt{\alpha_{em}}\,Q\,\alpha(1-\alpha)\,
\bar\chi\;\vec\sigma\cdot\vec n\;\chi\,
\Phi_0(\epsilon,r,\lambda)\ ,
\label{320}
 \EE
 where
 \BE
\lambda=
\frac{2\,a^2(\alpha)}{\epsilon^2}\ .
\label{330}
 \EE

The functions $\Phi_{0,1}$ in Eqs.~(\ref{310}) and (\ref{320})
are defined as
 \BE
\Phi_0(\epsilon,r,\lambda) =
\frac{1}{4\pi}\int\limits_{0}^{\infty}dt\,
\frac{\lambda}{{\rm sh}(\lambda t)}\,
{\rm exp}\left[-\ \frac{\lambda\epsilon^2 r^2}{4}\,
{\rm cth}(\lambda t) - t\right]\ ,
\label{340}
 \EE
 \BE
\vec\Phi_1(\epsilon,r,\lambda) =
\frac{\epsilon^2\vec r}{8\pi}\int\limits_{0}^{\infty}dt\,
\left[\frac{\lambda}{{\rm sh}(\lambda t)}\right]^2\,
{\rm exp}\left[-\ \frac{\lambda\epsilon^2 r^2}{4}\,
{\rm cth}(\lambda t) - t\right]\ .  
\label{350}
 \EE

Note that the $\bar q-q$ interaction enters Eqs.~(\ref{310}) and
(\ref{320}) via the parameter $\lambda$ defined in (\ref{330}). In the
limit of vanishing interaction $\lambda\to 0$ (i.e. $Q^2\to \infty$,
$\alpha$ is fixed, $\alpha\not=0$ or $1$) Eqs.~(\ref{310}) -
(\ref{320})  produce the perturbative expressions of Eq.~(\ref{70}).

With the choice $a^2(\alpha)\propto \alpha(1-\alpha)$ the end-point
behavior of the mean square interquark separation $\la r^2\ra\propto
1/\alpha(1-\alpha)$ contradicts the idea of confinement. Following
\cite{kst2} we fix this problem via a simple modification of the LC
potential,
 \BE
a^2(\alpha) = a^2_0 +4a_1^2\,\alpha(1-\alpha)\ .
\label{180}
 \EE 
 The parameters $a_0$ and $a_1$ were adjusted in \cite{kst2} to data on
total photoabsorption cross section \cite{gamma1,gamma2}, diffractive
photon dissociation and shadowing in nuclear photoabsorption reaction.  
The results of our calculations vary within only $1\%$ when $a_0$ and
$a_1$ satisfy the relation,
 \BA
a_0^2&=&v^{ 1.15}\, (0.112)^2\,\GeV^{2}\nonumber\\
a_1^2&=&(1-v)^{1.15}\,(0.165)^2\,\GeV^{2}\ , 
\label{190} 
 \EA
 where $v$ takes any value $0<v<1$. In view of this insensitivity of the
observables we fix the parameters at $v=1/2$. 
We checked that this choice does not affect our results beyond a
few percent uncertainty.

\subsection{The meson wave function}
\label{rho-wf}

To describe electroproduction reactions it is natural to work in the
infinite momentum frame of the virtual photon and use the LC variables
for the $\bar qq$ pair, the transverse separation $\vec r$ and the
fraction $\alpha=p_q^+/p_V^+$ of the total LC momentum carried by the
quark. The wave functions of light vector mesons are poorly known both in
the rest and infinite momentum frames. A popular prescription
\cite{terentev} is to apply the Lorentz boost to the rest frame wave
function assumed to be Gaussian which leads to radial parts of
transversely and longitudinally polarized mesons in the form,
 \BE
\Phi_V^{T,L}(\vec r,\alpha) = 
C^{T,L}\,\alpha(1-\alpha)\,f(\alpha)\,
{\rm exp}\left[-\ \frac{\alpha(1-\alpha)\,{\vec r}^2}
{2\,R^2}\right]\ 
\label{170}
 \EE
 with a normalization defined below, and 
 \beq
f(\alpha) = \exp\left[-\ \frac{m_q^2\,R^2}
{2\,\alpha(1-\alpha)}\right]\ .
\label{170a}
 \eeq 
 This procedure is ill motivated since the $\bar qq$ are not classical
particles. As a result of the boost to the infinite momentum frame many
new Fock components are created. Nevertheless, a detailed analysis of
this problem \cite{hz} leads to the same form (\ref{170}) which we use in
what follows with the parameters from \cite{jan97}, $R=0.59\,\fm$ and
$m_q=0.15\,\GeV$.

We assume that the distribution amplitude of $\bar qq$ fluctuations for
the vector meson and for the photon have a similar structure
\cite{jan97}.  Then in analogy to Eqs.~(\ref{310}) -- (\ref{320}),
 \beqn
\Psi^T_V(\vec r,\alpha) &=&
(A+\vec B\cdot\vec\nabla)\,
\Phi^T_V(r,\alpha)\ ;
\label{172}\\
\Psi^L_V(\vec r,\alpha) &=& 
2\,m_V\,\alpha(1-\alpha)\,
(\bar\chi\,\vec\sigma\cdot\vec n\,\chi)\,
\Phi^L_V(r,\alpha)\ .
\label{174}
 \eeqn

Correspondingly, the normalization conditions for the transverse 
and longitudinal vector meson wave
functions read,
 \BE
N_{C}\,\int d^{2} r\,\int d\alpha \left\{ m_{q}^{2}\,
\Bigl|\Phi^T_{V}(\vec r,\alpha)\Bigr|^{2} + \Bigl[\alpha^{2} + 
(1-\alpha)^{2}\Bigr]\,
\Bigl|\partial_{r}
\Phi^T_{V}(\vec r,\alpha)\Bigr|^{2} \right\} = 1
\label{230}
 \EE
 \BE
4\,N_{C}\,\int d^{2} r\,\int d\alpha\, 
\alpha^{2}\,(1-\alpha)^{2}\,
m_{V}^{2}\,\Bigl|\Phi^L_{V} (\vec r,\alpha)\Bigr|^2 = 1\, .
\label{240}
 \EE

\subsection{Cross section on a nucleon, comparison 
with data}\label{data-N}

Now we are in the position to calculate the forward production amplitude
$\gamma^*\,N \to V\,N$ for transverse and longitudinal photons and vector
mesons using the nonperturbative photon wave functions Eqs.~(\ref{310}),
(\ref{320}) and for the vector meson Eqs.~(\ref{172}), (\ref{174}).  We
verify the LC approach by comparing with data for nucleon target. This is
a rigorous test since we have no free parameters. 

The forward scattering amplitude reads,
 \BA
{\cal M}_{\gamma^{*}N\rightarrow V\,N}^{T}(s,Q^{2})
\Bigr|_{t=0} &=& 
N_{C}\,Z_{q}\,\sqrt{\alpha_{em}}
\int d^{2} r\,\sigma_{\bar qq}(\vec r,s)
\int\limits_0^1 d\alpha \Bigl\{ m_{q}^{2}\,
\Phi_{0}(\epsilon,\vec r,\lambda)\Psi^T_{V}(\vec r,\alpha)
\nonumber\\ 
&+& \bigl [\alpha^{2} + (1-\alpha)^{2}\bigr ]\,
\vec{\Phi}_{1}(\epsilon,\vec r,\lambda)\cdot
\vec{\nabla}_{r}\,\Psi^T_{V}(\vec r,\alpha) \Bigr\}\,;
\label{360}
 \EA
 \BA
{\cal M}_{\gamma^{*}N\rightarrow V\,N}^{L}(s,Q^{2})
\Bigr|_{t=0} &=& 
4\,N_{C}\,Z_{q}\,\sqrt{\alpha_{em}}\,m_{V}\,Q\,
\int d^{2} r\,\sigma_{\bar qq}(\vec r,s)\nonumber\\
&\times& \int\limits_0^1 d\alpha\,
\alpha^{2}\,(1-\alpha)^{2}\, 
\Phi_{0}(\epsilon,\vec r,\lambda)
\Psi^L_{V}(\vec r,\alpha)\ .
\label{370}
 \EA
 These amplitudes are normalized as ${|{\cal M}^{T,L}|^{2}}=
\left.16\pi\,{d\sigma_{N}^{T,L}/ dt}\right|_{t=0}$. We include the real
part of the amplitude according to the prescription described in
Sect.~\ref{dipole-cross}. In what follows we calculate the cross sections
$\sigma = \sigma^T + \epsilon\,\sigma^L$ assuming that the photon
polarization is $\epsilon=1$.

Now we can check the absolute value of the predicted cross section by
comparing with data for elastic electroproduction $\gamma^*\,p \to V\,p$
for $\rho$ and $\phi$ mesons.  Unfortunately, data are available only for
the cross section integrated over $t$, 
 \beq
\sigma^{T,L}(\gamma^{*}N\to VN) = 
\frac{|{\cal M}^{T,L}|^{2}}
{16\pi\,B_{\gamma^*N}}\ ,
\label{375}
 \eeq
 where the $t$-slope of the differential cross section which cannot be
properly predicted by the approach under consideration. Our strategy is
to predict the numerator in (\ref{375}), and compare with data for the
cross section and the slope.

\begin{figure}[tbh]
\includegraphics{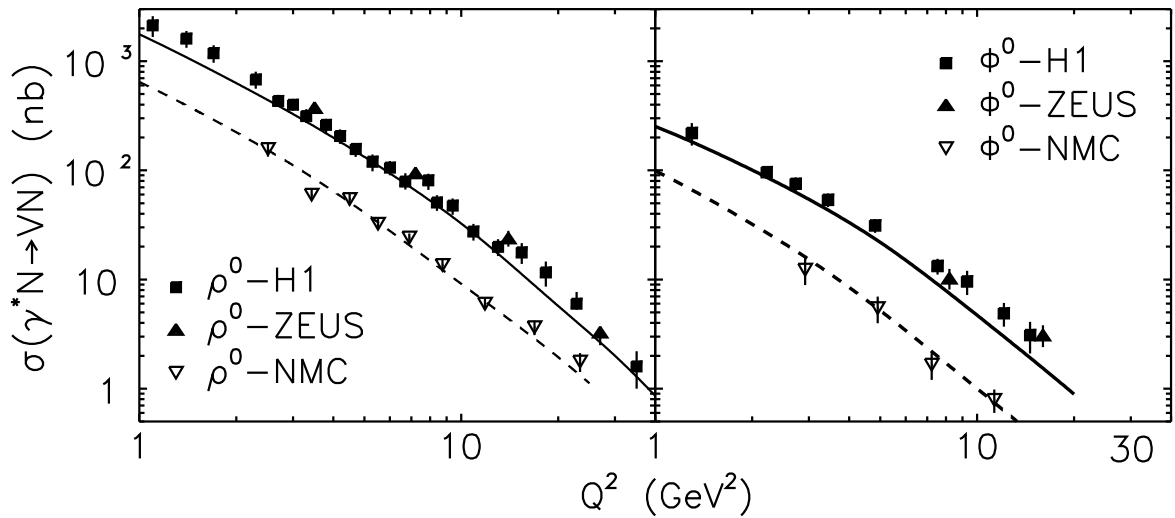}
\begin{center}
\vspace{7.2cm}
\parbox{13cm}
{\caption[Delta]
 {\sl $Q^2$- dependence of the cross section for the reactions
$\gamma^*\,p \to \rho\,p$ (left) and $\gamma^*\,p \to \phi\,p$ (right).  
The dashed and solid curves are compared with data at $W=15\,\GeV$
\cite{nmc-phirho-q2} and
at $75\,\GeV$ (\cite{H1-rho,ZEUS-rho} for $\rho$ and
\cite{q2-slope-phi-hera} for $\phi$), respectively.}
 \label{q2-nucl}}
\end{center}
 \end{figure}

Our predictions are plotted in Fig.~\ref{q2-nucl} together with the data
on the $Q^2$ dependence of the cross section from NMC, H1 and ZEUS 
\cite{nmc-phirho-q2,H1-rho,ZEUS-rho,q2-slope-phi-hera}).

 We use the $Q^2$ dependent slope of the differential cross section
$d\sigma(\gamma^*N\to VN)/dt \propto
\exp\Bigl[B^V_{\gamma^*N}(Q^2)\,t\Bigr]$ parametrized as \cite{jan98},
 \beq
B_{\gamma^*N}(s,Q^2) = \beta^V_0(s) + 
\frac{\beta^V_1(s)}{Q^2 + m_{V}^2} -
{1\over2}\,\ln\left(\frac{Q^2 + 
m_{V}^2}{m_{V}^2}\right)\ .
\label{slope-q2}
 \eeq
 A fit to the data \cite{nmc-phirho-q2,q2-slope-rho-lowenergy} from fixed
target experiments for the $Q^2$ dependent slope in $\rho$ production
give the parameters for $W\approx 10-15\,\GeV$,
$\beta^\rho_0=(6.2\pm0.2)\,\GeV^{-2}$, $\beta^\rho_1=1.5\pm0.2$. Using
data from HERA
\cite{H1-rho,ZEUS-rho,H1-slope-a,H1-slope-b,ZEUS-slope-a,ZEUS-slope-b,ZEUS-slope-c}
for $\rho$ production at $W=75\,\GeV$ we get,
$\beta^\rho_0=(7.1\pm0.1)\,\GeV^{-2}$, $\beta^\rho_1=2.0\pm0.1$.

Repeating the same analysis for $\phi$ production we get at $W\approx
10-15\,\GeV$ from data \cite{nmc-phirho-q2,real-phi-slope},
$\beta^\phi_0=(5.9\pm0.1)\,\GeV^{-2}$, $\beta^\phi_1=0.5\pm0.1$. Data
from HERA \cite{ZEUS-phi-0,q2-slope-phi-hera} give,
$\beta^\phi_0=(6.7\pm0.2)\,\GeV^{-2}$, $\beta^\phi_1=1.0\pm0.1$. For
calculations shown in Fig.~\ref{q2-nucl} we use the central values of
these parameters.

\begin{figure}[tbh]
\includegraphics{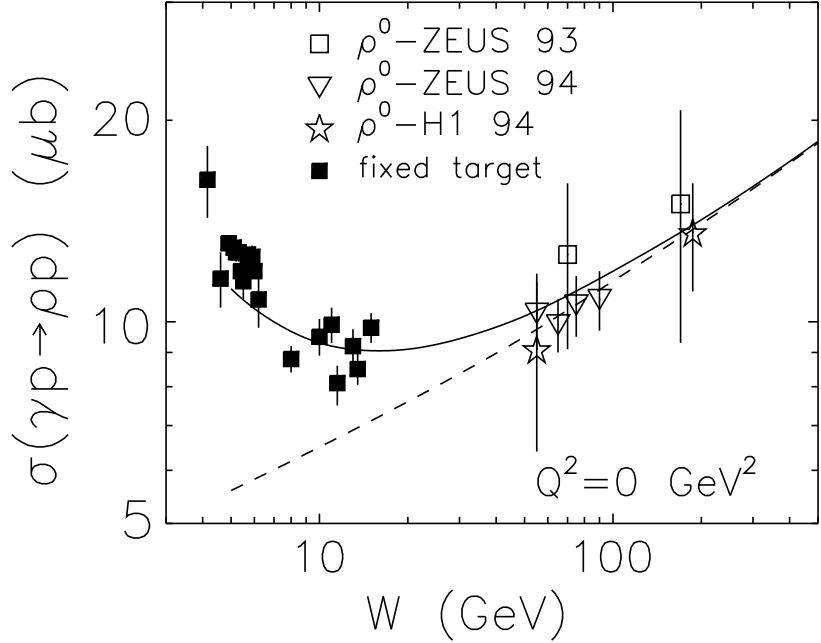}
\begin{center}
\vspace{9cm}
\parbox{13cm}
{\caption[Delta]
{\sl Energy dependence of the real photoproduction cross section
on a nucleon, $\gamma\,p \to \rho^0\,p$. Our results (solid curve) are 
compared with data from the fixed target 
\cite{le-rho-0a,le-rho-0b}, and collider HERA H1 \cite{H1-slope-b} and 
ZEUS \cite{ZEUS-slope-a,ZEUS-rho-0} experiments. The dashed
curve contains only the gluonic exchange in the t-channel.}
\label{real-nucl}}
\end{center}
 \end{figure}

Our approach which includes the effects of the nonperturbative
interaction between the $q$ and $\bar q$ in the photon fluctuation is
designed to describe the low $Q^2$ region as well. To test it we compare
with data \cite{le-rho-0a,le-rho-0b,H1-slope-b,ZEUS-slope-a,ZEUS-rho-0}
for the energy dependence of the cross section of real $\rho$
photoproduction in Fig.~\ref{real-nucl}.
 We use the energy dependent slope parameter, $B^\rho_{\gamma N}=B^\rho_0
+ 2\alpha'\ln(s/s_0)$ with $\alpha'=0.25\,\GeV^{-2}$ and
$B^\rho_0=7.6\,\GeV^{-2}$, $s_0=20\,\GeV^2$ fitted to data
\cite{real-rho-lowenergy,le-rho-0a,H1-slope-a,ZEUS-slope-a,ZEUS-slope-b}.
The Pomeron part of the dipole cross section depicted by the dashed curve
in Fig.~\ref{real-nucl} cannot explain the data at low energies, $W\lsim
15$\,GeV, while the addition of the Regge term (solid curve)  leads to a
good agreement for all energies. We also found a good agreement with data
for real photoproduction of $\phi$, but skip the comparison since there
are very few data points.

The normalization of the cross section and its energy and $Q^2$
dependence are remarkably well reproduced in Figs.~\ref{q2-nucl} --
\ref{real-nucl}. This is an important achievement since the absolute
normalization is usually much more difficult to reproduce than nuclear
effects. For instance, the similar, but simplified calculations in
\cite{kz-91} underestimate the $J/\Psi$ photoproduction cross section on
protons by an order of magnitude.

As a cross-check for the choice of the $\rho^{0}$ wave function in 
Eqs.~(\ref{170}) and (\ref{180}) we also calculated the total
$\rho^0$-nucleon cross section, which is usually expected to be roughly
similar to the pion-nucleon cross section $\sigma_{tot}^{\pi N}\sim 25$
mb. The $\rho$-nucleon total cross section has the form,
 \BA
\sigma_{tot}^{\rho N} = 
N_{C}\,\int d^{2} r\,
\int d\alpha \left\{ m_{q}^{2}\,
\Bigl|\Phi^T_{V} (\vec r,\alpha)\Bigr |^2 + 
\bigl [\alpha^{2} + (1-\alpha)^{2}\bigr ]\,
\Bigl |\partial_{r}\Phi^T_{V}(\vec r,\alpha)\Bigr|^{2} 
\right\}\,
\sigma_{\bar qq}(\vec r,s)
\label{380}
 \EA
 We calculated $\sigma_{tot}^{\rho N}$ with the $\rho$ meson wave
function in the form Eq.~(\ref{170}) with the parameters described in the
Sect.~\ref{rho-wf}. For the dipole cross section we adopt the KST
parameterization (\ref{130}) which is designed to describe low-$Q^2$
data. Then, at $\nu = 100\,\GeV$ we obtain $\sigma_{tot}^{\rho N} =
27\,\mb$ which is quite a reasonable number.

\section{Incoherent production of vector mesons off nuclei}\label{q-qbar}

In diffractive incoherent (quasielastic) production of vector mesons
off nuclei, $\gamma^{*}\,A\rightarrow V\,X$,
one sums over all final states of the target nucleus except those which 
contain particle (pion) creation. 
The observable usually studied experimentally is nuclear transparency 
defined as 
 \BE
Tr^{inc}_{A} = 
\frac{\sigma_{\gamma^{*}A\to VX}^{inc}}
{A\,\sigma_{\gamma^{*}N\to VN}}\ .
\label{480}
 \EE
 The $t$-slope of the differential quasielastic cross section is the same
as on a nucleon target. Therefore, instead of integrated cross sections
one can also use the forward differential cross sections Eq.~(\ref{125})
to write,
 \beq 
Tr^{inc}_A = \frac{1}{A}\,
\left|\frac{{\cal M}_{\gamma^{*}A\to VX}(s,Q^{2})}
{{\cal M}_{\gamma^{*}N\to VN}(s,Q^{2})}\right|^2\, .
\label{485}
 \eeq 

\subsection{The LC Green function approach}\label{green-f}

One should decompose the physical photon
$|\gamma^*\ra$ into different Fock states, namely, the bare photon
$|\gamma^*\ra_0$, $|\bar qq\ra$, $|\bar qqG\ra$, etc. The higher states
containing gluons are vital to describe the energy dependence of the
photoproduction reaction on a nucleon. As far as nuclear effects are
concerned, those Fock components also lead to gluon shadowing.  
However, as we mentioned above, these fluctuations are heavier and have a
shorter coherence time (lifetime) than the lowest $|\bar qq\ra$
state. Therefore, at medium energies only $|\bar qq\ra$ fluctuations of 
the photon matter. Gluon shadowing related to the higher Fock
states will be considered later.

Propagation of an interacting $\bar qq$ pair in a nuclear medium is
described by the Green function satisfying the evolution Eq.~(\ref{250}).
However, the potential in this case acquires an imaginary part which
represents absorption in the medium (see (\ref{10}) for notations),
 \BE
Im V_{\bar qq}(z_2,\vec r,\alpha) = - 
\frac{\sigma_{\bar qq}(\vec r,s)}{2}\,\rho_{A}({b},z_2)\,.
\label{440}
 \EE
 The evolution equation (\ref{250}) with the potential
$V_{\bar qq}(z_{2},\vec r_{2},\alpha)$ containing this imaginary 
part
was used in \cite{krt1,krt2}, and nuclear shadowing in deep-inelastic
scattering was calculated in good agreement with data.

The analytical solution of Eq.~(\ref{270}) is only known for the 
harmonic oscillator potential $V(r)\propto r^2$. To keep
the calculations reasonably simple we are forced to use the dipole 
approximation,
 \beq
\sigma_{\bar qq}(r,s) = C(s)\,r^2\ .
\label{460}
 \eeq
 The energy dependent factor $C(s)$ is adjusted to reproduce correctly
nuclear effects in the limit of very long CL $l_c\gg R_A$ (the so
called ``frozen'' approximation), when 
 \beq
G_{\bar qq}(z_1,\vec r_1;z_2,\vec r_2) \Rightarrow
\delta(\vec r_1-\vec r_2)\,\exp\left[
-{1\over2}\,\sigma_{\bar qq}(r_1)
\int\limits_{z_1}^{z_2} dz\,\rho_A(b,z)\right]\ ,
\label{465}
 \eeq 
 where the dependence of the Green function on impact parameter is
dropped. The details are described in Appendix A.

With the potential Eqs.~(\ref{440}) -- (\ref{460}) the solution of 
Eq.~(\ref{250}) has the same form as Eq.~(\ref{270}), except that one 
should 
replace $\omega \Rightarrow \Omega$, where
 \beq
\Omega = \frac{\sqrt{a^4(\alpha)-
i\,\rho_{A}({b},z)\,
\nu\,\alpha\,(1-\alpha)\,C(s)}}
{\nu\;\alpha(1-\alpha)}\ .
\label{470}
 \eeq

Guided by the uncertainty principle and the Lorentz transformation one
can estimate the coherence time as in Eq.~(\ref{30}), where the effective
mass of the $\bar qq$ pair is replaced by the vector meson mass. One can
see the presence of a coherence length in the kinetic term of the
evolution equation Eq.~(\ref{250}). Indeed, the effective mass squared of
a $\bar qq$ pair is $M^2_{\bar qq}=(m_q^2+k_T^2)/\alpha(1-\alpha)$. This
is what the kinetic term consists of when the transverse momentum squared
of the quark is replaced by $k_T^2 \Rightarrow \Delta_{r}$. This
dynamically varying effective mass controls the CL defined by the Green
function, as compared to the oversimplified Eq.~(\ref{30}) for the CL as
given by the fixed mass $m_V$. One can explicitly see the static part
$Q^2+m_q^2/\alpha(1-\alpha)$ of the coherence length in the last phase
shift factor in the Green function in Eq.~(\ref{270}).

Depending on the value of $l_c$ one can distinguish different regimes:

{\bf (i)} The CL is much shorter than the mean nucleon spacing in a
nucleus ($l_c \to 0$). In this case $G(z_2,\vec r_2;z_1,\vec r_1)  \to
\delta(z_2-z_1)$ since strong oscillations suppress propagation of the
$\bar qq$ over longer distances. In this case the formation time of the
meson wave function is very short as well, since it is described by the
same Green function and is controlled by the formation time as given in
Eq.~(\ref{20}). Apparently, for light vector mesons $l_f\sim l_c$, so
both must be short. In this case nuclear transparency is given by the
simple formula Eq.~(\ref{40}) corresponding to the Glauber
approximation\footnote{Note that the optical approximation is used
throughout this paper only for the sake of easy reading. For numerical
calculations we replace the exponential by a more realistic expression,
$\exp(-\sigma\,T_A) \Rightarrow (1-\sigma\,T_A/A)^{A-1}$. This has also
been done in all of our previous publications, contrary to what is
stated in \cite{wolfram}.}.

\begin{figure}[tbh]
\includegraphics{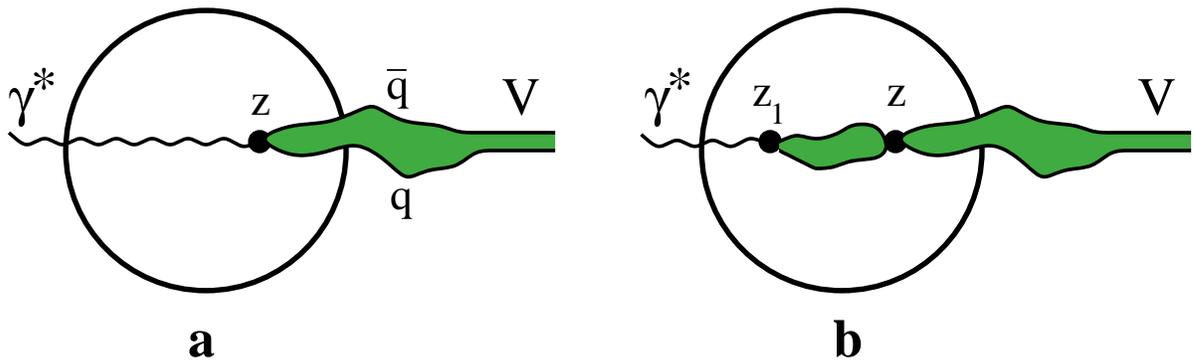}
\begin{center}
\vspace{5.3cm}
\parbox{13cm}
 {\caption[Delta]
 {\sl The incident virtual photon produces incoherently at the point $z$
(quasielastic scattering)  the colorless $\bar qq$ pair which then
evolves propagating through the nucleus and forms the V-meson wave
function ({\bf a}). Alternatively, the photon can first produce
diffractively and coherently at the point $z_1$ the colorless $\bar qq$
which then experiences quasielastic scattering at the point $z$ ({\bf
b}). Propagation of the $\bar qq$ pair is described by the Green function
(shaded areas).}
 \label{pict}}
\end{center}
 \end{figure}

{\bf (ii)}
In the intermediate case $l_c\to 0$, but $l_f\sim R_A$, which can  
only be realized for heavy flavor quarkonia, the formation of the meson 
wave function 
is described by the Green function and the numerator of the nuclear 
transparency ratio Eq.~(\ref{485}) has the form 
\cite{kz-91},
 \beq
\Bigl|{\cal M}_{\gamma^{*}A\to VX}(s,Q^{2})
\Bigr|^2_{l_c\to0;\,l_f\sim R_A} = 
\int d^2b\int_{-\infty}^{\infty} dz\,\rho_A(b,z)\,
\Bigl|F_1(b,z)\Bigr|^2\ ,
\label{500}
 \eeq
 where
 \beq
F_1(b,z) = 
\int_0^1 d\alpha
\int d^{2} r_{1}\,d^{2} r_{2}\,
\Psi^{*}_{V}(\vec r_{2},\alpha)\,
G(z^\prime,\vec r_{2};z,\vec r_{1})\,
\sigma_{\bar qq}(r_{1},s)\,
\Psi_{\gamma^{*}}(\vec r_{1},
\alpha)\Bigl|_{z^\prime\to\infty}
\label{505}
 \eeq
This expression is illustrated in Fig.~\ref{pict}a.
 The photon creates at the point $z$ a colorless $\bar qq$ pair with
transverse separation $\vec r_1$. The quark and antiquark then
propagate through the nucleus along different trajectories and end up
with a separation $\vec r_2$. The contributions from different paths
are summed up giving rise to the Green function $G(z^\prime,\vec
r_2;z,\vec r_1)$ which is convoluted in (\ref{500}) with
the wave functions of $\gamma^*$ and $V$. This is the path integral
technique suggested in \cite{kz-91}.

{\bf (iii)} $l_c \gg R_A$ (in fact, it is more correct to compare with
the mean free path of the $\bar qq$ in a nuclear medium if the latter is 
shorter
than the nuclear radius). In this case $G(z_2,\vec r_2;z_1,\vec r_1)
\to \delta(\vec r_2 - \vec r_1)$, i.e. all fluctuations of 
the transverse $\bar qq$ 
separation are ``frozen'' by Lorentz time dilation.
Then, the numerator on the r.h.s. of Eq.~(\ref{485}) takes the form
\cite{kz-91},
 \beqn
\Bigl|{\cal M}_{\gamma^{*}A\to VX}(s,Q^{2})
\Bigr|^2_{l_c \gg R_A} &=& 
\int d^2b\,T_A(b)\left|\int d^2r\int_0^1 d\alpha
\right. \label{510}\\
&\times& \left.\Psi_{V}^{*}(\vec r,\alpha)\,   
\sigma_{\bar qq}(r,s)\,  
\exp\left[-{1\over2}\sigma_{\bar qq}(r,s)\,T_A(b)\right]
\Psi_{\gamma^{*}}(\vec r,\alpha,Q^2)\right|^2\ .
\nonumber
 \eeqn 
 In this case the $\bar qq$ attenuates with a constant absorption cross
section like in the Glauber model, except that the whole exponential is
averaged rather than just the cross section in the exponent. The
difference between the results of the two prescriptions are the well 
known inelastic corrections of Gribov \cite{zkl}.

{\bf (iv)} The main and new results of the present paper address the
general case with no restrictions for either $l_c$ or $l_f$.  No
theoretical tool has been developed so far beyond the limits (i) -- (iii)
discussed above neither of which can be applied to electroproduction of
light vector mesons at the medium high energies of HERMES and JLab.

Even within the VDM the Glauber model expression interpolating between
the limiting cases of low [(i), (ii)] and high [(iii)] energies has been
derived only recently \cite{hkn}. We generalize that formalism to the LC
dipole approach, and the incoherent photoproduction amplitude is
represented as a sum of two terms \cite{hkz} illustrated in
Fig.~\ref{pict},
 \BE 
\Bigl|\,{\cal M}_{\gamma^{*}A\to
VX}(s,Q^{2})\Bigr|^{2} = \int d^{2}b
\int\limits_{-\infty}^{\infty} dz\,\rho_{A}({b},z)\, 
\Bigl|F_{1}({b},z) - F_{2}({b},z)\Bigr|^{2}\ .
\label{520}
 \EE
 The first term $F_{1}({b},z)$ introduced above in Eq.~(\ref{505}) is
represented by Fig.~\ref{pict}a. Alone it would correspond to the short
$l_c$ limit (ii). The second term $F_{2}({b},z)$ in (\ref{520})  
corresponds to the situation illustrated in Fig.~\ref{pict}b. The
incident photon produces a $\bar qq$ pair diffractively and coherently at
the point $z_1$ prior to incoherent quasielastic scattering at point $z$.
The LC Green functions describe the evolution of the $\bar qq$ over the
distance from $z_1$ to $z$ and further on, up to the formation of the
meson wave function. Correspondingly, this term has the form,
 \beqn
F_{2}(b,z) &=& \frac{1}{2}\,
\int\limits_{-\infty}^{z} dz_{1}\,\rho_{A}(b,z_1)\,
\int\limits_0^1 d\alpha\int d^2 r_1\,
d^2 r_{2}\,d^2 r\,
\Psi^*_V (\vec r_2,\alpha)\nonumber \\
&\times&
G(z^{\prime}\to\infty,\vec r_2;z,\vec r)\,
\sigma_{\bar qq}(\vec r,s)\,
G(z,\vec r;z_1,\vec r_1)\,
\sigma_{\bar qq}(\vec r_1,s)\,
\Psi_{\gamma^{*}}(\vec r_1,\alpha)\, .
\label{530}
 \eeqn

Eq.~(\ref{520}) correctly reproduces the limits (i) - (iii). Indeed, at
$l_c\to 0$ the second term $F_2(b,z)$ vanishes because of strong
oscillations, and Eq.~(\ref{520}) reproduces the Glauber expression
Eq.~(\ref{40}). On the other hand, at $l_c\gg R_A$ the phase shift in the
Green functions can be neglected and they acquire the simple form
$G(z_2,\vec r_2;z_1,\vec r_1) \to \delta(\vec r_2 - \vec r_1)$. In this
case the integration over longitudinal coordinates in Eqs.~(\ref{505})
and (\ref{530}) can be performed explicitly and the asymptotic expression
Eq.~(\ref{510}) is recovered as well. Moreover, if one uses a constant
dipole cross section $\sigma_{\bar qq}(\rho)=\sigma^{VN}_{tot}$, then
Eq.~(\ref{520}) recovers the general Glauber expression\footnote{Note
that Eq.~(\ref{520}) and its Glauber model analog in \cite{hkn} include
all coherent multiple scattering terms, contrary to a statement made in
\cite{wolfram}.} derived in \cite{hkn}.

\subsection{Data for incoherent production: CT or
coherence?}\label{incoh-data}

Exclusive incoherent electroproduction of vector mesons off nuclei has
been suggested in \cite{knnz} as a sensitive way to detect CT. Increasing
the photon virtuality $Q^2$ one squeezes the produced $\bar qq$ wave
packet. Such a small colorless system propagates through the nucleus with
little attenuation, provided that the energy is sufficiently high
($l_f\gg R_A$) the fluctuations of the $\bar qq$ separation are frozen by
Lorentz time dilation. Thus, a rise of nuclear transparency
$Tr_A^{inc}(Q^2)$ with $Q^2$ should signal CT\footnote{This process has a
definite advantage compared to quasielastic electron scattering (e,e'p)  
suggested in \cite{mueller,brodsky} as a probe for CT. Indeed, in the
latter case the energy of the photon correlates with its virtuality,
$\nu\approx 2m_NQ^2$, and one has to increase $Q^2$ just in order to
increase $\nu$ and keep the size of the ejectile ``frozen''. This leads
to a substantially diminished cross section, which is why no CT signal
has been detected in this reaction so far (it is still possible to
observe CT in this reaction at low energy studying the asymmetry of the
quasielastic peak as function of $x_{Bj}$ \cite{jk}). In contrast, no
correlation between $\nu$ and $Q^2$ exists in exclusive electroproduction
of vector mesons.}.  Indeed, such a rise was observed in the E665
experiment at Fermilab for exclusive production of $\rho^0$ mesons off
nuclei by a muon beam. This has been claimed in \cite{e665-rho} to be a
manifestation of CT.

However, one should be cautious to avoid mixing up the expected signal
for CT with the effect of coherence length \cite{kn95,hkn}. Indeed, if
the coherence length varies from long to short compared to the nuclear
size the nuclear transparency rises because the length of the path in
nuclear matter becomes shorter and the vector meson (or $\bar qq$)
attenuates less. This happens when $Q^2$ increases at fixed $\nu$. One
should carefully disentangle these two phenomena.

\begin{figure}[tbh]
\includegraphics{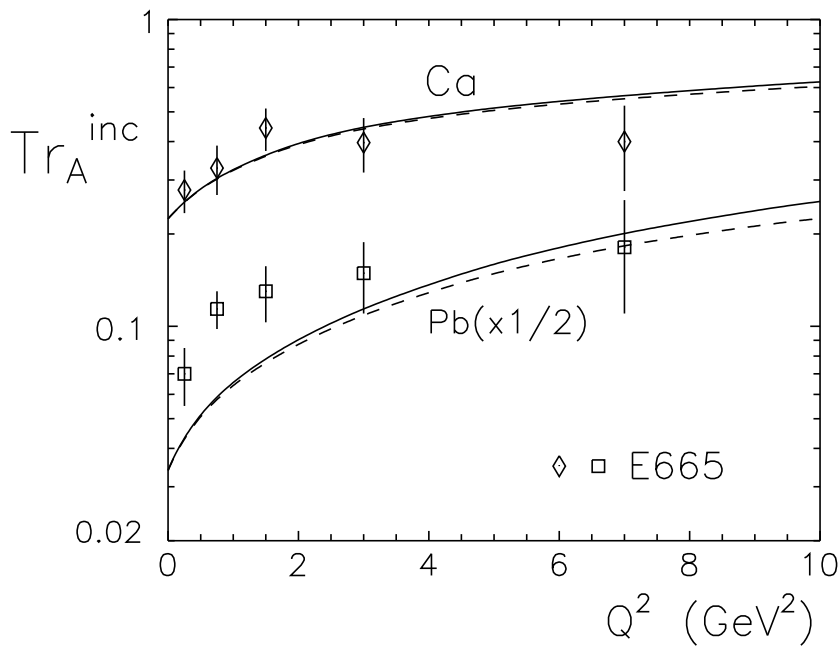}
\begin{center}
\vspace{8.6cm}
\parbox{13cm}
{\caption[Delta]
{\sl 
$Q^2$- dependence of nuclear transparency for lead and calcium 
$Tr_{Pb}$ and $Tr_{Ca}$ vs. 
The experimental points are from the E665 experiment 
\cite{e665-rho}. Both the curves and data for lead are rescaled by the 
factor $1/2$.
Solid and dashed curves show our results using the LC Green function 
approach Eq.~(\ref{520}) and the ``frozen'' approximation  
Eq.~(\ref{510}) respectively.}
\label{e665}}
\end{center}
 \end{figure}

{\bf Long CL.} It has been checked in \cite{kn95} that the coherence
length at the kinematics of the E665 experiment is sufficiently long to
neglect its variation with $Q^2$ and to use the ``frozen'' approximation,
except at the highest values of $Q^2\gsim 5\,\GeV^2$.  We calculated
nuclear transparency, $Tr_A^{inc}$, of incoherent (quasielastic)
$\rho^{0}$ production using Eq.~(\ref{520}) and the simplified ``frozen''
approximation Eqs.~(\ref{465}) -- (\ref{510}). The results are depicted
in Fig.~\ref{e665} by solid and dashed curves respectively.
 One can see that fluctuations of the size of the $\bar qq$ pair become
important only at high $Q^2$ causing a separation of the solid and dashed
curves. At smaller $Q^2$ the observed variation of $Tr_A^{inc}(Q^2)$ is a net
manifestation of CT. The agreement with our model is surprisingly good 
for calcium, while we underestimate the nuclear transparency at small 
$Q^2$ for lead.
This may be a manifestation of large Coulomb corrections as found in
\cite{ktv}, which are of the order $\alpha_{em}Z\approx 0.6$ for lead.
These corrections lead to a considerable deviation from the Born,
one-photon approximation employed in \cite{e665-rho} in order to obtain data
for $\gamma^*\,A\to\rho^0\,X$ (depicted in Fig.~\ref{e665}) from raw data
for $\mu\,A\to\mu'\,\rho^0\,X$. This important problem needs further study.

{\bf Medium long CL.} The same process of incoherent electroproduction of
$\rho^0$ is under study at lower energies, in the HERMES experiment at
HERA and at JLab. In this case one should carefully discriminate between
the effects of CT and CL \cite{kn95,hkn}. A simple prescription
\cite{hk-97} to eliminate the effect of CL from the data on the $Q^2$
dependence of nuclear transparency is to bin the data in a way which
keeps $l_c = const$. It means that one should vary simultaneously $\nu$
and $Q^2$ maintaining the CL Eq.~(\ref{30}) constant,
 \beq
\nu = {1\over2}\,l_c\,(Q^2+m_V^2)\ .
\label{534}
 \eeq
 In this case the Glauber model predicts a $Q^2$ independent nuclear
transparency, and any rise with $Q^2$ would signal CT \cite{hk-97}.

\begin{figure}[tbh] 
\includegraphics{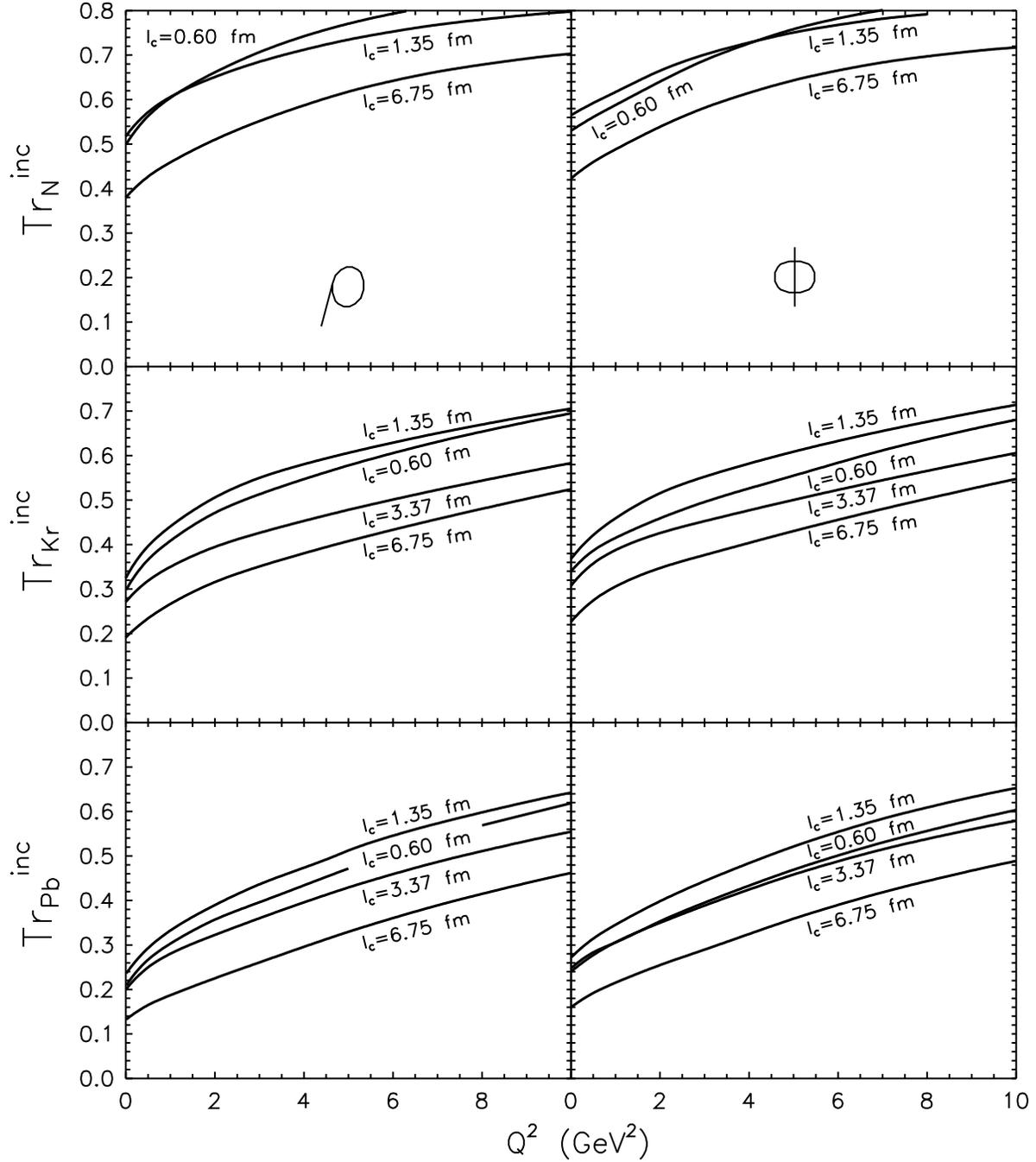} 
\begin{center}
\vspace{17cm} 
\parbox{13cm} 
{\caption[Delta] 
 {\sl $Q^2$ dependence of the nuclear transparency $Tr_A^{inc}$ for
exclusive electroproduction of $\rho$ (left) and $\phi$ (right)  mesons
on nuclear targets $^{14}N$, $^{84}Kr$ and $^{207}Pb$ (from top to
bottom). The CL is fixed at $l_c = 0.60$, $1.35$, $3.37$ and
$6.75$\,fm.}
 \label{lc-const-inc}}
\end{center}
 \end{figure}

The LC Green function technique incorporates both the effects of
coherence and formation. We performed calculations of $Tr_A^{inc}(Q^2)$
at fixed $l_c$ starting from different minimal values of $\nu$, which
correspond to real photoproduction in Eq.~(\ref{534}),
 \beq
\nu_{min}={1\over2}\,l_c\,m_V^2\ . 
\label{536} 
 \eeq
The results for incoherent production of $\rho$ and $\phi$ at 
$\nu_{min}= 0.9,\ 2,\ 5$ and $10\,\GeV$ ($l_c=0.6 - 
6.75\,\fm$) are presented in Fig.~\ref{lc-const-inc} for nitrogen, 
krypton and lead. 
 We use the nonperturbative LC wave function of the photon with the
parameters of the LC potential $a_{0,1}$ fixed in accordance with
Eq.~(\ref{190}) at $v=1/2$. The $u$ and $d$ quarks are assumed to be
massless, but we use $m_s=0.15\,\GeV$. Nuclear transparency for $\phi$ is
stronger than for $\rho$ as one could expect, but the difference is not
significant. In what follows we discuss only our results for $\rho$.

For $\rho$-mesons the predicted variation of nuclear transparency with
$Q^2$ at fixed $l_c$ is much stronger than was found in \cite{hk-97}.
Those calculations have been done in the hadronic representation which is
quite challenging due to the necessity to know all the diagonal and off
diagonal diffractive amplitudes for the vector meson and its excitations,
as well as all photoproduction amplitudes. The predictions made in
\cite{hk-97} were based on the two-coupled-channel model without any
estimate of the accuracy of such an approximation. According to
quark-hadron duality the LC Green function method is equivalent to the
exact solution of the general multi-channel problem in the hadronic
representation. The comparison therefore demonstrates that the
two-channel approximation substantially underestimates the effect of
color transparency.

\begin{figure}[tbh]
\includegraphics{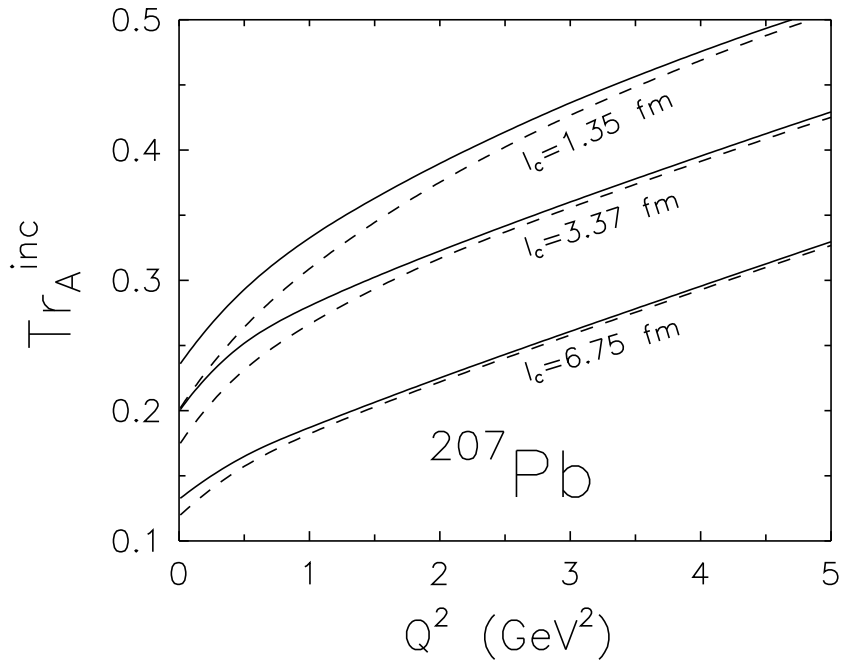}
\begin{center}
\vspace{9cm}
\parbox{13cm}
{\caption[Delta]
{\sl The same as in Fig.~\ref{lc-const-inc} for lead, but calculated
with both nonperturbative (solid curves) and perturbative (dashed) wave 
functions of the photon.}
\label{npt-effects}}
\end{center}
 \end{figure}

 To see the scale of the theoretical uncertainty of our model \cite{kst2}
for nonperturbative effects we compare in Fig.~\ref{npt-effects} the
results for the $\rho$-meson obtained using the nonperturbative (solid
curves)  and perturbative photon wave functions Eq.~(\ref{70}) with
$m_q=0.15\,\GeV$ (dashed curves). The difference between the two sets of
curves is insignificant.

Motivated by the too weak signal predicted for CT it was suggested in
\cite{hk-97} that instead one can study the effect of coherence which has
never been observed experimentally. Indeed, it was found in
\cite{lc-hermes} that data for nuclear transparency for $\rho$ production
plotted as function of $l_c$ well agree with what was predicted in
\cite{hkn} to be the effect of the CL. Now we find a rather strong signal
of CT which may also affect the $l_c$ dependence of $Tr_A^{inc}$ and
cause a deviation from the Glauber model expectations. We therefore
revise the previous conclusions \cite{hk-97,lc-hermes}.

\begin{figure}[htb]
\includegraphics{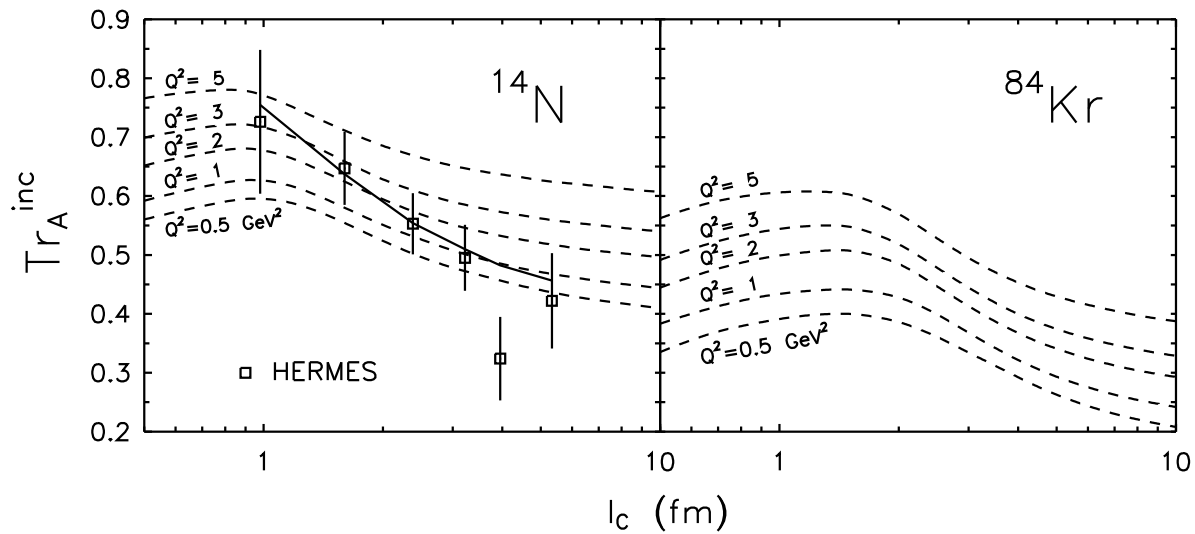}
\begin{center}
\vspace{8cm}
\parbox{13cm}
{\caption[Delta]
 {\sl Nuclear transparency for incoherent electroproduction of $\rho$ off
nuclei, nitrogen and krypton, as function of $l_c$ at fixed $Q^2=0.5,\
1,\ 2,\ 3,\ 5\,\GeV^2$. The solid curve is calculated at the mean values
of $l_c$ and $Q^2$ corresponding to each experimental point
\cite{lc-hermes,borissov}.}
 \label{lc-hermes-fig}}
\end{center}
 \end{figure}

In the VDM-Glauber model nuclear transparency is a function of $l_c$ only
(neglecting the weak energy dependence of $\sigma^{VN}_{tot}$), however
it becomes a function of two variables, $Tr_A^{inc}(l_c,Q^2)$, as soon as
CT effects are involved. Therefore, our current predictions for the $l_c$
dependence of $Tr_A^{inc}$ vary with $Q^2$. They are plotted by dashed
curves in Fig.~\ref{lc-hermes-fig} for different fixed values of
$Q^2=0.5,\ 1,\ 2,\ 3,\ 5\,\GeV^2$ (from bottom to top) for nitrogen and
krypton (left and right boxes, respectively).
 The nonperturbative wave function of the photon was used as for
Fig.~\ref{lc-const-inc}. We do not show the results obtained with the
perturbative wave function since they are pretty much the same, except in
the region of small $Q^2$ and short $l_c$ where they are about $10\%$
lower that the nonperturbative results.

The experimental points for nitrogen \cite{lc-hermes} which are plotted
in Fig.~\ref{lc-hermes-fig} correspond to different mean experimental
values of $Q^2$ \cite{borissov}. This $Q^2 - l_c$ correlation is
incorporated in our calculations, and the results depicted by the solid
curve agree well with the data.

We thus arrive at the conclusion that the two quite different approaches,
the VDM based Glauber model and QCD based LC Green function formalism,
both provide good agreement with the HERMES data. This could not be
possible if the data were plotted as function of $l_c$ at fixed $Q^2$.
The observed agreement with the Glauber model seems to be accidental and
a result of the $Q^2 - l_c$ correlation in the data.

In order to discriminate between the two approaches one should plot the
data differently.  Fig.~\ref{lc-hermes-fig} gives hope that the data are
sufficiently accurate to detect a signal of CT if they are properly
analyzed. Also additional data for krypton should soon become available
from HERMES.

The expected signal for CT is a nonzero derivative $d\,\ln[Tr_A
(Q^2)]/d\,Q^2$ which is predicted in Fig.~\ref{lc-const-inc} to be
similar for different nuclei and different values of $l_c$. One can make
use of this fact and perform a common fit to all available data with only
one parameter which is the slope of the $Q^2$ dependence of nuclear
transparency. The value of the logarithmic slope for the mid values, $Q^2
\approx 1 - 2\, \GeV^2$, of the HERMES kinematical 
range for $\rho$ production is expected to vary within the
interval
 \beq
\frac{1}{Tr^{inc}_A (Q^2)}\ 
\frac{d\,Tr^{inc}_A (Q^2)}{d\,Q^2}\Bigr|_{l_c=const}
\approx
\begin{array}{l}
0.07 - 0.11\ \GeV^{-2}\ \ {\rm for}\ \  ^{14}N\\
0.14 - 0.17\ \GeV^{-2}\ \ {\rm for}\ \  ^{84}Kr
\end{array}
\label{derivative}
 \eeq
for $l_c = 0.60 - 6.75\ \fm$. Similar, but somewhat smaller values of the 
logarithmic $Q^2$-slope are expected for $\phi$.

The curves in Fig.~\ref{lc-const-inc} demonstrate an interesting
property. The slope of the $Q^2$ dependence is steeper at small $Q^2$ and
$l_c$. For instance, the logarithmic derivative Eq.~(\ref{derivative})
equals $0.09$ at $l_c=0.6\,\fm$, but is smaller, $0.07$ at
$l_c=1.35\,\fm$ This fact might be in variance with naive intuitive
expectations. Indeed, $l_c=0.6\,\fm$ is short compared to th mean spacing
of the bound nucleons. Since $l_f\sim l_c$ at low $Q^2$ one might expect
the Glauber model to be a good approximation in this case. Apparently,
this is not the case, Fig.~\ref{lc-const-inc} demonstrates a steepest
growth of $Tr_A^{inc}(Q^2)$ in this region. One can understand this as
follows. If $l_c$ is long, like in Fig.~\ref{e665}, then the formation
length is long too, $l_f\gsim l_c\gg R_A$, and nuclear transparency rises
with $Q^2$ only because the mean transverse separation of the $\bar qq$
fluctuations decreases. If, however, $l_c\lsim R_A$ and fixed, the photon
energy rises with $Q^2$ according to Eq.~(\ref{534}) and the formation
length Eq.~(\ref{20}) rises as well. Thus, these two effects, the $Q^2$
dependence of $l_f$ and the $\bar qq$ transverse size, add up and lead to
a steeper growth of $Tr_A^{inc}(Q^2)$ for short $l_c$.

One should conclude from this consideration that the CT effects are more
pronounced at low than at high energies. This observation adds to the
motivation for experimental searches for CT at HERMES and JLab.

\begin{figure}[htb]
\includegraphics{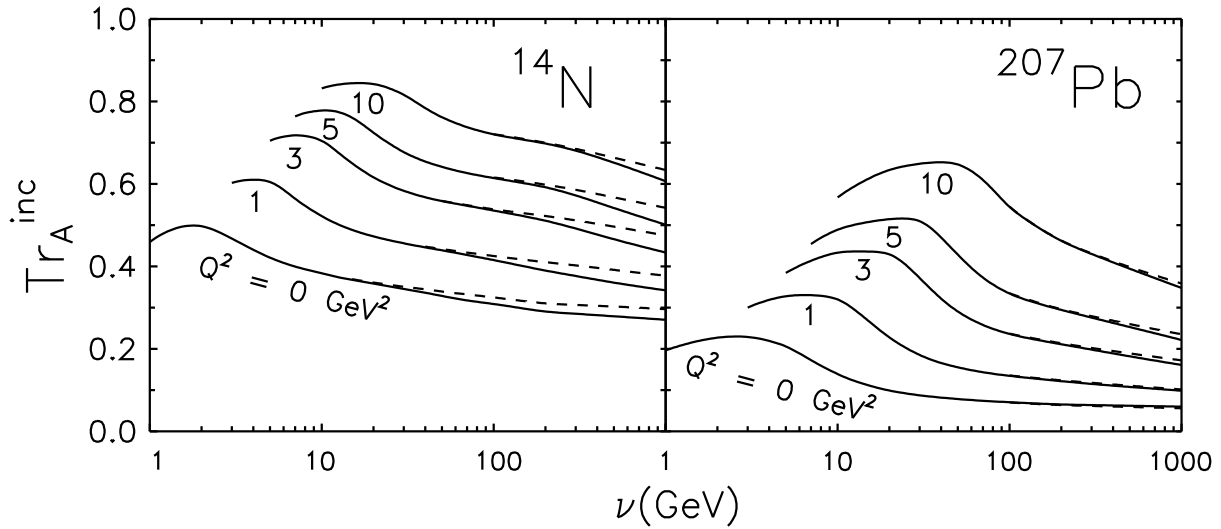}
\begin{center}
\vspace{7.2cm}
\parbox{13cm}
{\caption[Delta]
 {\sl Nuclear transparency for incoherent electroproduction $\gamma^*A\to
\rho^0 A$ as function of energy at $Q^2=0,\ 1,\ 3,\ 5,\ 10\,\GeV^2$ for
nitrogen and lead. The solid and dashed curves correspond to calculations
with and without gluons shadowing, respectively.}
 \label{glue-incoh}}
\end{center}
 \end{figure}

We also calculated the energy dependence of nuclear transparency at fixed 
$Q^2$. The results for nitrogen and lead are shown by dashed curves in 
Fig.~\ref{glue-incoh} for different values of $Q^2$.
 The interesting feature is the presence of a maximum of transparency at
some energy. It results from the interplay of coherence and formation
effects. Indeed, the FL rises with energy leading to an increasing
nuclear transparency. At some energy, however, the effect of CL which is
shorter than the FL, is switched on leading to a growth of the path
length of the $\bar qq$ in the nucleus, i.e. to a suppression of
transparency. The maxima in the $l_c$ dependence of nuclear transparency
depicted in Fig.~\ref{lc-hermes-fig} are of the same nature.  This also
explains the unusual ordering of curves calculated for different values
of $l_c$ as is depicted in Fig.~\ref{lc-const-inc}.

\section{Coherent production of vector mesons}\label{coh}

\subsection{The formalism}\label{coh-form}

If electroproduction of a vector meson leaves the target intact the
process is usually called coherent or elastic.  The mesons produced at
different longitudinal coordinates and impact parameters add up
coherently. This fact considerably simplifies the expressions for the
cross sections compared to the case of incoherent production. The
integrated cross section has the form,
 \BE
\sigma_A^{coh}\equiv
\sigma_{\gamma^{*}A\to VA}^{coh} = 
\int d^2q\,\left|\int d^2b\,
e^{i\vec q\cdot\vec b}\,
{\cal M}_{\gamma^{*}A\to VA}^{coh}(b)
\right|^2 = 
\int d^{2}\,{b}\,
|{\cal M}_{\gamma^{*}A\to VA}^{coh}
({b})\,|^{2}\ ,
\label{550}
 \EE
 where
 \BE
{\cal M}_{\gamma^{*}A\to VA}^{coh}({b}) =
\int\limits_{-\infty}^{\infty}\,dz\,\rho_{A}({b},z)\,
F_{1}({b},z)\ ,
\label{560}
 \EE
 with the function $F_{1}({b},z)$ defined in (\ref{505}).

One should not use Eq.~(\ref{485}) for nuclear transparency any more
since the $t$-slopes of the differential cross sections for nucleon and
nuclear targets are different and do not cancel in the ratio. Therefore,
the nuclear transparency also includes the slope parameter
$B_{\gamma^*N}$ for the process $\gamma^{*}\,N\rightarrow V\,N$,
 \BE
Tr_{A}^{coh} = \frac{\sigma_{A}^{coh}}{A\,\sigma_{N}} = 
\frac{16\,\pi\,B_{\gamma^*N}\,\sigma_{A}^{coh}}{A\,
|{\cal M}_{\gamma^{*}N\to VN}(s,Q^{2})\,|^{2}}
\label{570}
 \EE

One can also define a $t$-dependent transparency for coherent 
electroproduction of vector mesons,
 \BE
Tr_{A}^{coh}(t) =
\frac{ d\sigma_{A}^{coh} / dt }
{A^{2}\,d\sigma_{N} / dt |_{t=0}}\ ,
\label{600}
 \EE
 where the differential cross section for coherent production
$\gamma^{*}\,A\rightarrow V\,A$ reads
 \BE
\frac{d\sigma_{A}^{coh}}{dt} = 
\frac{1}{16\,\pi}\, 
\left|\int d^{2}{b}
\,e^{i\vec b\cdot\vec q}\,
\int\limits_{-\infty}^{\infty} 
dz\,\rho_{A}({b},z)\,F_{1}({b},z)\,
\right|^{2}\,
\label{610}
 \EE
 with $F_{1}({b},z)$ defined in (\ref{505}). This expression is
simplified in the limit of long coherence time ($t=-q^2$),
 \beqn
\frac{d\sigma_{A}^{coh}}{dt}
\Bigr|_{l_c\gg R_A} &=& 
\frac{1}{4\,\pi}\, 
\left|\int d^2 b\,
e^{i\vec b\cdot\vec q}\int d^2 r\,
\Biggl\{1\ -\ \exp\left[-\ {1\over2}\,
\sigma_{\bar qq}(\vec r,s)\,T(b)\right]\Biggr\}
\right.\nonumber\\ &\times& \left.
\int\limits_0^1 d\alpha\,
\Psi^{*}_{V}(\vec r,\alpha)\,
\Psi_{\gamma^{*}}(\vec r,\alpha)
\right|^2\ ,
\label{615}
 \eeqn
 a form which resembles its VDM analogue \cite{bauer}. 

\subsection{Comparison with data and predictions for coherent production}
\label{coh-data}

\begin{figure}[tbh]
\includegraphics{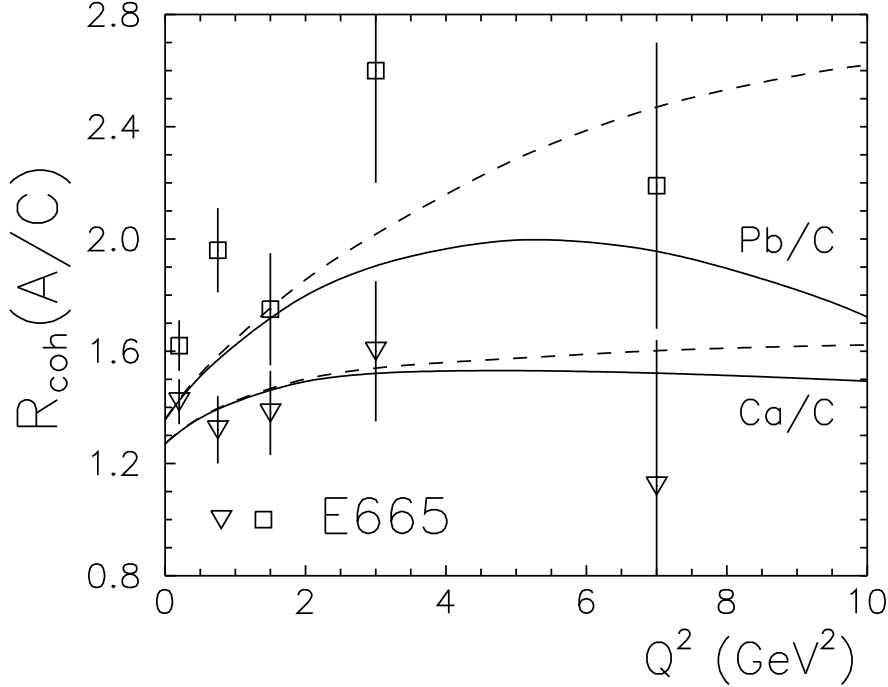}
\begin{center}
\vspace{9.5cm}
\parbox{13cm}
{\caption[Delta]
 {\sl $Q^{2}$- dependence of the total cross section ratio $R_{coh} (A/C)
= 12\sigma_A^{coh} / A\sigma_C^{coh}$ for the coherent process $\gamma^*
A\to \rho^0 A$. Experimental points are from E665 
\cite{e665-rho} for $Pb/C$ (squares)  and $Ca/C$ (triangles). Solid
curves include the variation of $l_c$ and $l_f$ with $Q^2$.
Dashed curves are calculated in the ``frozen'' approximation $l_c\gg
R_A$.}
 \label{e665-coh}}
\end{center}
 \end{figure}

Using Eq.~(\ref{570}) we can also calculate the normalized ratio of
coherent cross sections on two nuclei $R_{coh}(A_1/A_2) =
Tr_{A_1}^{coh}/Tr_{A_2}^{coh}$. The results of calculations for $R_{coh}(Pb/C)$ and
$R_{coh}(Ca/C)$ are depicted by solid curves in Fig.~\ref{e665-coh} as
well as corresponding data from the E665 experiment \cite{e665-rho} shown by
squares and triangles, respectively.
 We performed calculations of $Tr_A^{coh}$ at mean photon energy $\bar\nu
= 138\,\GeV$ with the $Q^2$ dependent slope given by
Eq.~(\ref{slope-q2}). All effects of CL and CT are included via the LC
Green function formalism.  For such a high energy one can think that the
``frozen'' approximation $l_c\gg R_A$ is good.  In order to check how
variation of the CL affects the nuclear transparency we repeated our
calculations in the ``frozen'' approximation and plotted the results as
dashed curves in Fig.~\ref{e665-coh}. We see that the accuracy of this
approximation is rather good for calcium, while for lead it significantly
deviates from the exact result at $Q^2\gsim 2\,\GeV^2$. The reason is
obvious, the heavier the nucleus, the less the approximation $l_c\gg R_A$
is fulfilled. We also see that the contraction of the CL with $Q^2$
causes an effect opposite to CT, namely nuclear transparency is
suppressed rather than enhanced. Therefore, there is no danger that CL
effects can mock CT, and one may think that this is an advantage of
coherent compared to incoherent production \cite{kn95}. However, at
medium energy the suppression of nuclear transparency at short CL is so
strong that no rise of nuclear transparency with $Q^2$ might be
observable.

Note that in contrast to incoherent production where nuclear transparency
is expected to saturate as $Tr^{inc}_A(Q^2) \to 1$ at large $Q^2$, for
the coherent process nuclear transparency reaches a higher limit,
$Tr^{coh}_A(Q^2) \to A^{1/3}$ (of course, $A^{1/3}$ is valid only for
very large nuclei, otherwise it is an approximate number). The dashed
curves in Fig.~\ref{e665-coh} nearly reach this upper limit at $Q^2\sim
10\,\GeV^2$.

\begin{figure}[tbh] 
\includegraphics{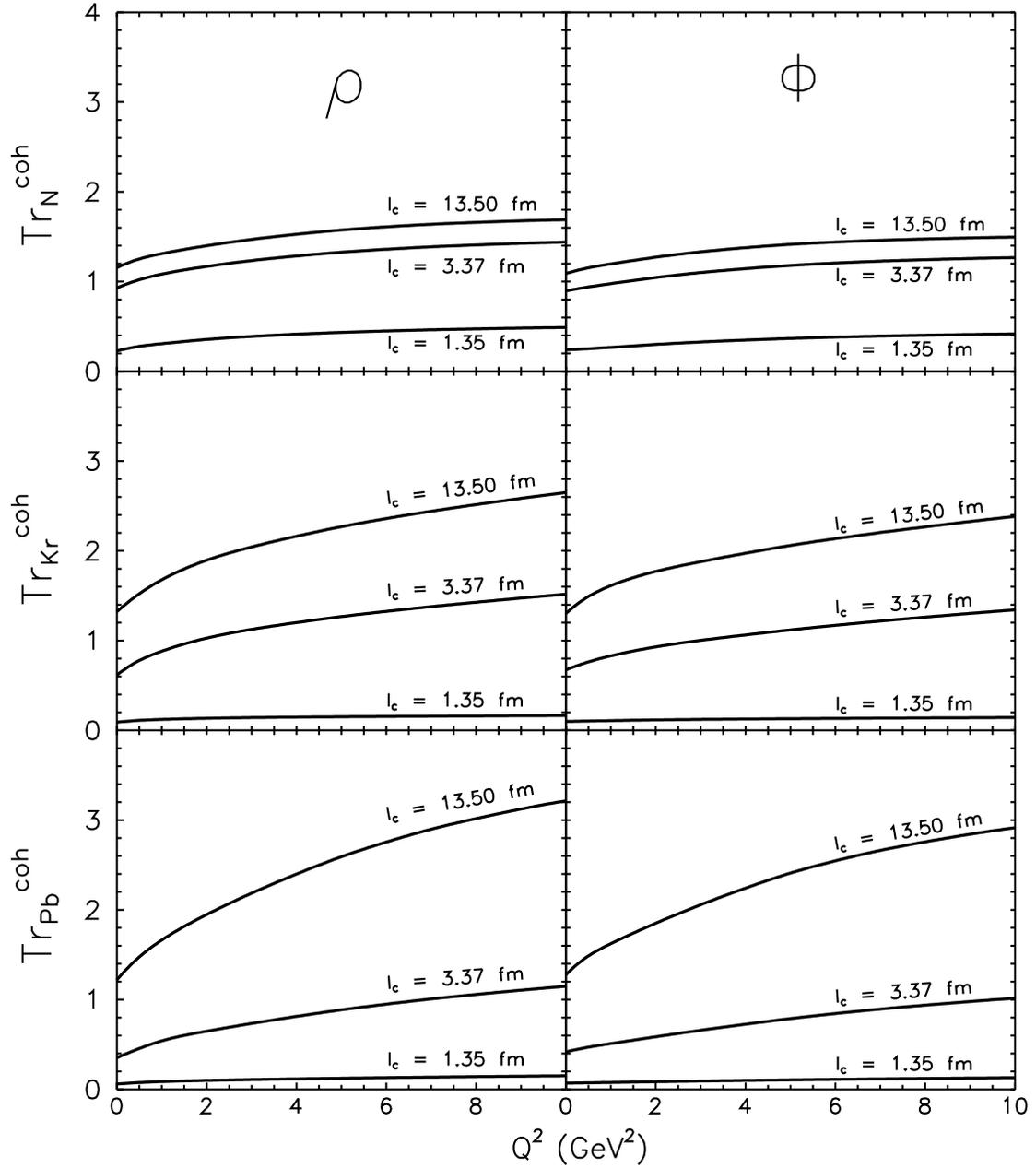} 
\begin{center}
\vspace{17cm} 
\parbox{13.3cm} 
{\caption[Delta] 
 {\sl The same as in Fig.~\ref{lc-const-inc}, but for coherent production
of $\rho$ and $\phi$, $\gamma^*A\to V A$.}
 \label{lc-const-coh}}
\end{center}
 \end{figure}

One can eliminate the effects of CL and single out the net CT effect in a
way similar to what was suggested for incoherent reactions by selecting
experimental events with $l_c=const$. We calculated nuclear transparency
for the coherent reaction $\gamma^*A\to \rho(\phi) A$ at fixed values of
$l_c$. The results for $l_c=1.35,\ 3.37,\ 13.50\,\fm$ are depicted in
Fig.~\ref{lc-const-coh} for several nuclei.
 The effect is sufficiently large to be observable, the logarithmic 
derivative varies within the interval,
 \beq
\frac{1}{Tr^{coh}_A (Q^2)}\ 
\frac{d\,Tr^{coh}_A (Q^2)}{d\,Q^2}\Bigr|_{l_c=const}
\approx
\begin{array}{l}
0.14 - 0.07\ \GeV^{-2}\ \ {\rm for}\ \  ^{14}N\\
0.10 - 0.15\ \GeV^{-2}\ \ {\rm for}\ \  ^{84}Kr
\end{array}
\label{derivative-coh}
 \eeq
 for $l_c = 1.35 - 13.5\ \fm$. Again, like in the case of incoherent
production, the logarithmic derivative decreases at large $l_c$.  
The magnitude of the expected CT effect is similar to the value predicted
for the incoherent production in Eq.~(\ref{derivative}) and is slightly 
smaller for $phi$ than for $\rho$.

\begin{figure}[htb]
\includegraphics{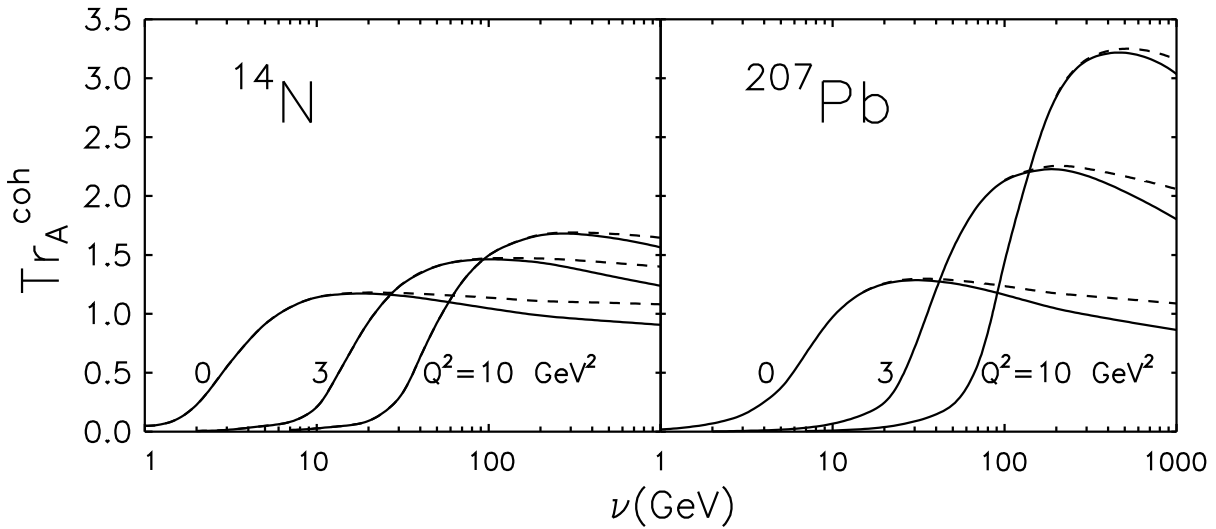}
\begin{center}
\vspace{7.2cm}
\parbox{13cm}
{\caption[Delta]
 {\sl Nuclear transparency for incoherent electroproduction 
$\gamma^*A\to \rho A$ as function of energy at $Q^2=0,\ 3,\ 10\,\GeV^2$
for nitrogen and lead. The solid and dashed curves correspond to 
calculations with and without gluon shadowing, respectively.}
 \label{glue-coh}}
\end{center}
 \end{figure}

We also  calculated nuclear transparency as function of energy at fixed 
$Q^2$. The results for $\rho$ produced coherently off nitrogen and lead 
are depicted by dashed curves
in Fig.~\ref{glue-coh} at $Q^2=0,\ 3,\ 10\,\GeV^2$. 
 $Tr^{coh}_A$ is very small at low energy, what of course does not mean
that nuclear matter is not transparent, but the nuclear coherent cross
section is suppressed by the nuclear form factor. Indeed, the
longitudinal momentum transfer which is equal to the inverse CL, is large
when the CL is short. However, at high energy $l_c\gg R_A$ and nuclear
transparency nearly saturates (it decreases with $\nu$ only due to the
rising dipole cross section). The saturation level is higher at larger
$Q^2$ which is a manifestation of CT.

\subsection{Transverse momentum distribution}\label{t-dep}

\begin{figure}[htb]
\includegraphics{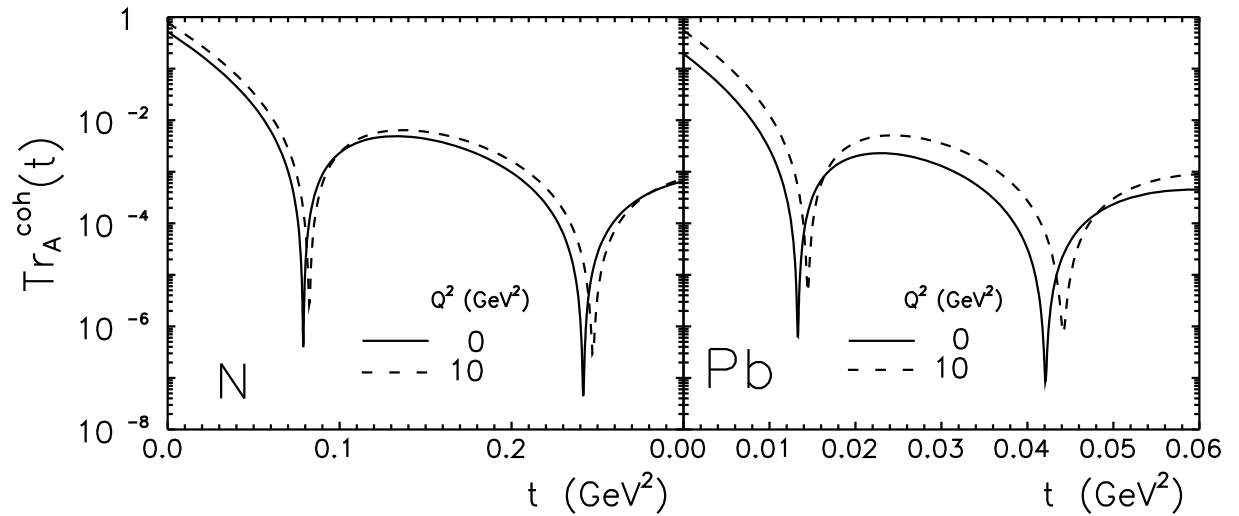}
\begin{center}
\vspace{7cm}
\parbox{13cm}
{\caption[Delta]
 {\sl Nuclear transparency for coherent electroproduction 
$\gamma^*A\to \rho^0 A$ as function of momentum transfer squared
calculated for nitrogen and lead in the limit of $l_c\gg R_A$.
The solid and dashed curves correspond to $Q^2=0$ and $10\,\GeV^2$ 
respectively.}
 \label{t-dep-coh}}
\end{center}
 \end{figure}

Another manifestation of CT is a modification of the diffractive pattern
in the momentum transfer dependence of the coherent cross section
\cite{wolfram}. Indeed, the effect of CT on the nuclear transparency
depends on impact parameter. Fourier transformation of such a modified
amplitude will apparently result in a shifted positions of the
diffractive minima. Indeed, calculations performed in the ``frozen''
approximation assuming sufficiently high energy ($l_c\gg R_A$) lead to
the $t$-dependence of nuclear transparency from Eq.~(\ref{600}) depicted
in Fig.~\ref{t-dep-coh}.
 We see that the CT effects shift the position of the diffractive minima
to larger $t$. To understand the sign of the effect, we can use the
approximate dipole cross section $\sigma_{\bar qq}(r) = C\,r^2$. Further,
we can approximate the product of the photon and vector meson wave
functions by a Gaussian $\propto \exp(-r^2/\la r^2\ra)$. The partial
amplitude (amplitude for given impact parameter) of elastic production
$\gamma^* \to \rho^0$ takes the form \cite{zkl},
 \beq
\int d^2 r\,\Psi^*_{V}(\vec r)\,
\Psi_{\gamma^*}(\vec r,Q^2)\,
\left\{1-\exp\left[-{1\over2}\,
C\,r^2\,T_A(b)\right]\right\} =
\frac{C\,\la r^2\ra\,T_A(b)}
{2 + C\,\la r^2\ra\,T_A(b)}\ .
\label{1000}
 \eeq 

The mean size $\la r^2\ra$ of the $\bar qq$ wave packet decreases with
$Q^2$ suppressing the partial amplitude from Eq.~(\ref{1000}). It follows
from (\ref{1000}) the suppression is smaller on the periphery of the
nucleus than in the center. This implies that the slope of the
$t$-distribution $B_{\gamma^* A}=\la b^2\ra/2$ should decrease with
$Q^2$, i.e. the minima should move to larger values of $t$.

Although such a modification of the diffractive pattern should signal CT,
the effect is very weak and its observation does not look feasible.

\section{Other pitfalls in the search for CT at low
energies}\label{pitfalls}

In order to avoid the effect of CL which leads for incoherent production
to a nuclear transparency rising with $Q^2$ and mimics CT we suggested in
Sect.~\ref{incoh-data} to study the $Q^2$ dependence in data samples
which are preselected to have the same $l_c$. As soon as the CL effects
are excluded, the Glauber model predicts no variation of the transparency
with $Q^2$. There are, however, still other effects, not related to CT,
which cause a growth of CT even if $l_c=const$.

\subsection{Standard inelastic corrections}\label{standard-corr}

The effect of CT can be treated in the hadronic representation as a
multichannel problem \cite{zkl,jm,hk-97}: the incident virtual photon
produces diffractively on a bound nucleon either the ground state $V$, or
any excitation. Only this stage of the process is $Q^2$ dependent. The
produced states propagate further through the nucleus experiencing
diagonal and off diagonal diffractive transitions. Eventually, the state
$V$ is detected at macroscopic distances. These modifications of the
Glauber single-channel approximation are at the heart of Gribov's
inelastic shadowing.

The miracle of CT is the expectation that all those large amplitudes must
cancel leaving only one amplitude, namely the direct production of the
$V$. There is no hint from the hadronic representation that this should
happen. We have no data for most of those amplitudes and no hope to
measure them in future. Only the gauge invariance of QCD dictates this
very nontrivial behavior which is not present in any of the old fashion
models (colorless constituent quarks, etc.).

\begin{figure}[tbh]
\includegraphics{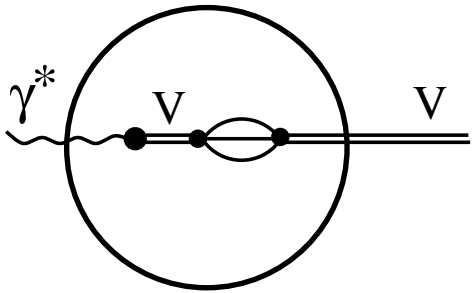}
\begin{center}
\vspace{5.3cm}
\parbox{13cm}
 {\caption[Delta]
 {\sl Intermediate diffractive excitation of a vector meson 
produced by the incident virtual photon and propagating 
through the nucleus.}
 \label{kk-pict}}
\end{center}
 \end{figure}

Nevertheless, we do have data for single diffraction which allow to
calculate some of the lowest order inelastic corrections. Although these
corrections are part of the whole CT phenomenon they are model
independent (provided that those models are fitted to available data). In
particular, the nuclear medium is known to be more transparent  
than expected using the Glauber model \cite{zkl}. Indeed, if the produced
$V$ state experiences inelastic diffraction inside the nucleus, it is
gone from the detected channel according to the rules of the Glauber
approximation. However, there is still a possibility to recover and come
back to the $V$ channel in a subsequent collision, as is illustrated in
Fig.~\ref{kk-pict}.
 Apparently, this process increases the survival probability for the $V$
state. There is clear experimental evidence that this takes place. The
total cross sections of hadron- (neutrons, neutral kaons) nucleus
interactions measured with high accuracy \cite{murthy,gsponer} are
smaller than the Glauber model predictions. This deviation increases with
energy as it is controlled by the nuclear form factor which depends on
the longitudinal momentum transfer $q_L$ in the diffraction dissociation.

Therefore, transparency of nuclear matter for hadrons increases with
energy and this fact leads to a rising $Q^2$ dependence if data are
selected according to the condition $l_c=const$. Indeed, the energy rises
according to the correlation $\nu=(Q^2+m_V^2)/2l_c$.  Of course this
effect is well known for the total cross sections since the pre-QCD era.
It cannot be (and never was) interpreted as a manifestation of CT.
Although these inelastic corrections are part of the CT phenomenon, one
should admit that they exist independently of the answer to the question
whether the CT is true or not.

Thus, one should be cautious in interpreting a rise of nuclear
transparency as function of $Q^2$ at fixed $l_c$.  The correction under
discussion calculated in \cite{kn} for quasielastic high-$p_T$ electron
scattering, $A(e,e'p)X$ was found to be indistinguishable from the
predicted CT effect up to rather high $Q^2$ of a few tens of $\GeV^2$.

The deviation of the transparency from the Glauber model prediction is
calculated as \cite{kn},
 \beqn
Tr_A^{inc}(Q^2) &=& \int d^2b\int\limits_{-\infty}^{\infty}
dz\,\rho_A(b,z)\,\exp\left[-\ \sigma^{VN}_{in}
\int\limits_z^\infty dz'\,\rho_A(b,z')\right]
\nonumber\\ &\times&
\left[1\ +\ 4\pi\int dM^2\,
\left.\frac{d\sigma(VN\to XN)}{dM^2\,dt}\right|_{t=0}\,
F_A(b,z,q_L)\right]^2\ .
\label{kk}
 \eeqn 
 Here $F_A(b,z,q_L)$ in (\ref{kk}) is the so called longitudinal
form factor of the nucleus calculated at a given impact parameter $b$ and
production coordinate $z$,
 \beq 
F_A(b,z,q_L) = \int\limits_{z}^{\infty}
dz'\,\rho_A(b,z')\,\cos(q_Lz')\ .
\label{formfactor}
 \eeq

\begin{figure}[tbh] 
\includegraphics{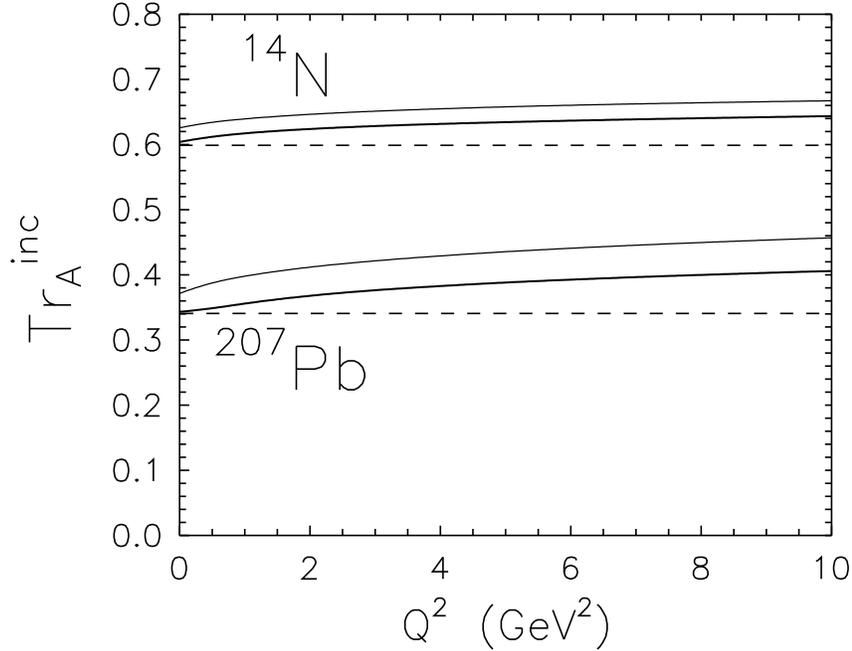} 
\begin{center} 
\vspace{9cm}
\parbox{13cm} 
{\caption[Delta] 
{\sl The same as in Fig.~\ref{lc-const-inc}, but
without any CT effects. Only the standard inelastic corrections which 
make
the nuclear matter more transparent are included. The dashed curves
correspond to the Glauber approximation. Each couple of solid curves
correspond to $l_c=1.35\,\fm$ (bottom curve) and $l_c=6.75\,\fm$
(upper curve). The two sets of curves correspond to nitrogen (top) and
lead (bottom)} 
\label{in-corr}} 
\end{center}
 \end{figure}

 We use the same parameterization for the single diffraction cross
section as in \cite{murthy,kn} except for the normalization which is
reduced by the factor $2/3$ as is suggested by the triple Regge
phenomenology. Although it is a rather rough estimate, it is sufficient
for our purpose, since the effect turns out to be very weak. The results
of our calculations for the $Q^2$ dependence of the transparency are
depicted by solid curves in Fig.~\ref{in-corr} at
 fixed $l_c= 1.35$ and $6.75\,\fm$ (the bottom and upper curves
respectively). One can see that although $Tr_A^{inc}$ grows with $Q^2$,
this effect when compared with Fig.~\ref{lc-const-inc} is too weak to be
mixed up with the signal of CT. Indeed, the derivative
$d\,\ln(Tr_A^{inc})/d\,Q^2$ evaluated at $Q^2=1 - 2\,\GeV^2$ equals
$0.011$ for nitrogen, $0.025$ for krypton and $0.033$ for lead. This is
nearly an order of magnitude less than what was estimated in
(\ref{derivative}) as a signal for CT.

\subsection{Finiteness of the \boldmath$\rho$ meson
lifetime}\label{lifetime}

Some of the effects have been calculated above at rather low energies
when the lifetime of the $\rho$ is comparable with the nuclear size.  
For instance, at $\nu=2\,\GeV$ the mean path length up to the decay is
only $2.7\,\fm$. Two pions have a smaller survival probability than the
$\rho$ therefore the nuclear transparency should be smaller than is
expected within the Glauber approximation disregarding decays. However,
as function of energy the decay path length increases and eventually the
nuclear transparency must reach the value corresponding to the Glauber
model. Again, at $l_c=const$ the energy of $\rho$ rises with $Q^2$ and
the nuclear transparency must grow with the decay length. This effect
might cause a problem in identifying the signal of CT at low energy, and
it should be considered with care. This correction is of less importance
for $\phi$ production.

We corrected the Glauber formula for the finite decay length in the 
following way,
 \beqn
Tr_A^{coh} &=& \gamma(\nu)\,
\int d^2b\int\limits_{-\infty}^{\infty}
dz_1\,\rho_A(b,z)\int\limits_{z_1}^{\infty} dz_2
\nonumber\\ &\times&
\exp\left[-\ \sigma^{\rho N}_{in}
\int\limits_{z_1}^{z_2} dz'\,\rho_A(b,z')\ 
-\ 2\,\sigma^{\pi N}_{in}
\int\limits_{z_2}^{\infty} dz'\,\rho_A(b,z')\ 
+\ \gamma(\nu)(z_2-z_1)\right]\ ,
\label{decay}
 \eeqn
 where 
 \beq
\gamma(\nu) = \frac{\Gamma_{\rho}\,m_{\rho}}
{\sqrt{\nu^2-m_{\rho}^2}}\ ,
\label{gamma}
 \eeq
 is the Lorentz enhanced decay length of the $\rho$ meson,
$\Gamma_{\rho}=0.15\,\GeV$ is the total decay width of the $\rho$. Like
previously, $\nu$ correlates with $Q^2$ via Eq.~(\ref{534}).

\begin{figure}[tbh] 
\includegraphics{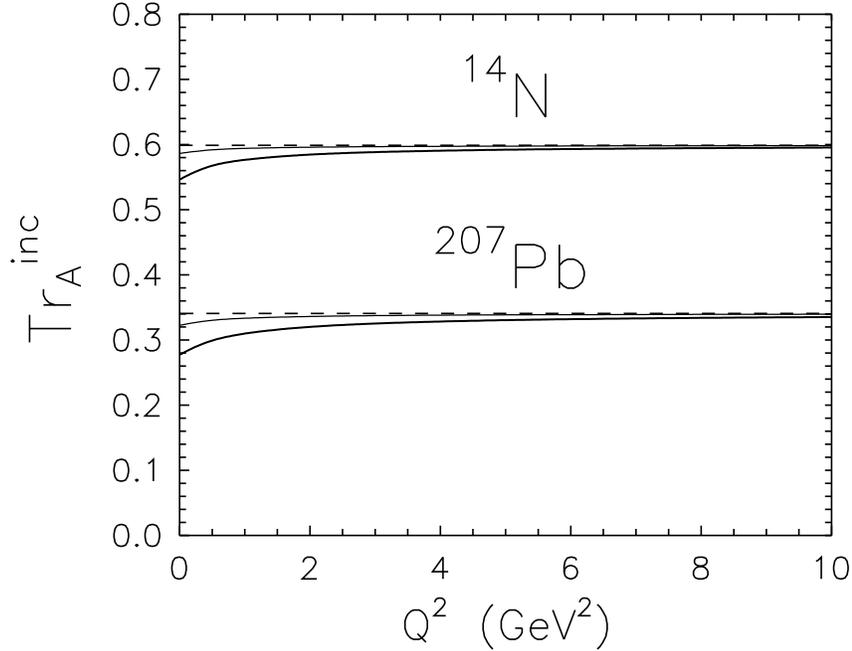} 
\begin{center} 
\vspace{9cm}
\parbox{13cm} 
{\caption[Delta] 
 {\sl The same as in Figs.~\ref{lc-const-inc} and \ref{in-corr}, but with
no effects of either CT or any inelastic correction included.  Only the
finite $\rho$ meson decay length which leads to an enhanced absorption is
taken into account. For each nucleus the dashed curve corresponds to the
Glauber approximation ($\Gamma_{\rho}\to 0$), the bottom and upper solid
curves correspond to $l_c=1.35$ and $6.75\,\fm$ respectively.}
 \label{decay-q2}} 
\end{center}
 \end{figure}

The results of our calculations are shown in Fig.~\ref{decay-q2}
as function of $Q^2$ at different fixed $l_c$.
 The effect turns out to be very weak compared to the expected effect of
CT demonstrated in Fig.~\ref{lc-const-inc}. For a CL of $l_c=1.35\,\fm$
($6.75\,\fm$) the results are shown by the bottom (upper) solid curves
for both nuclei. The derivative $d\,\ln(Tr_A^{inc})/d\,Q^2$ evaluated at
$Q^2=1 - 2\,\GeV^2$ equals $0.014$ ($0.003$) and $0.03$ ($0.008$) for
nitrogen and lead respectively. We conclude that the effect of the finite
$\rho$ decay length cannot be mixed up with a CT signal.

It is disputable whether the two pions emerging from a $\rho$ decay
immediately starts to attenuate with twice the pion absorption cross
section. One might think about two pions which strongly overlap at the
production point, then their cross section should be reduced due to color
screening. This is a general problem of how one should decide whether
decay has already happened or not. It is easier to understand for the
example of photon radiation by an electron. One can treat the photon as
being originally a part of the static Coulomb field of the electron which
is then shaken off by the interaction with a target. Only when the photon
and electron become incoherent they start acting as independent partons.
It takes, however, a time span dictated by the uncertainty principle to
discriminate between a coherent system, electron and its field, and an
incoherent pair of an electron plus a photon. This time is called
radiation or coherence time. In analogy, one can say that the $\rho$
meson has already decayed when the two pions become incoherent. In this
case they interact independently and Eq.~(\ref{decay}) is valid. However,
while the pions are still coherent, one should treat them as intrinsic
components of the $\rho$ meson.

\section{Gluon shadowing}\label{glue-shadow}

At very small $x_{Bj}$ the density of gluons should eventually deviate
from the predicted by the DGLAP evolution (as we mentioned, whether an
indication of this saturation effect was already seen at HERA is
controversial). In the infinite momentum frame this phenomenon looks like
gluon-gluon fusion. It corresponds to a nonlinear term in the evolution
equation \cite{glr}.  This effect should lead to a suppression of
small-$x_{Bj}$ gluons in a nucleon.  

One may expect a precocious onset of the saturation effects for heavy
nuclei.  In the infinite momentum frame of the nucleus the gluon clouds
of nucleons which have the same impact parameter overlap at small
$x_{Bj}$ in longitudinal direction. This allows gluons originated from
different nucleons to fuse leading to a gluon density which is not
proportional to the density of nucleons any more. This is gluon
shadowing.

Such a parton model interpretation is not Lorentz invariant, the same
phenomenon looks quite different in the rest frame of the nucleus. It
corresponds to the process of gluon radiation and shadowing corrections
related to multiple interactions of the radiated gluons in the nuclear
medium \cite{al}. This is a coherence phenomenon known as the
Landau-Pomeranchuk effect, namely the suppression of bremsstrahlung by
interference of radiation from different scattering centers. It demands
a sufficiently long coherence time of radiation, a condition equivalent
to demanding a small Bjorken $x_{Bj}$ in the parton model.

\begin{figure}[tbh]
\includegraphics{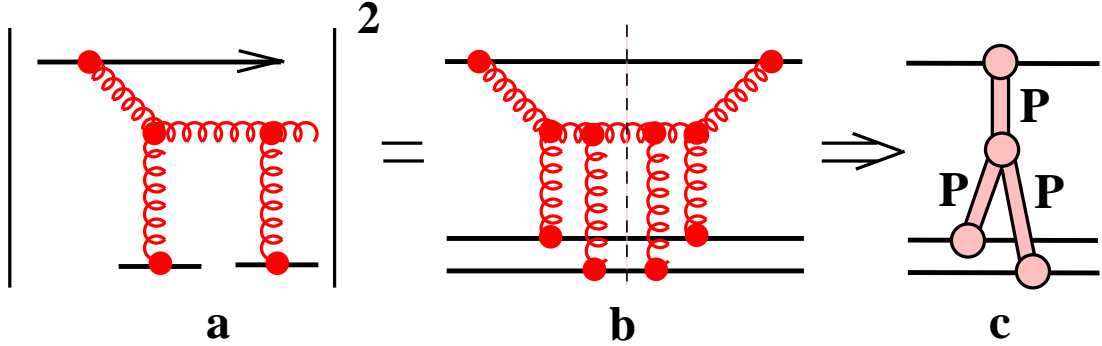}
\begin{center}
\vspace{4.5cm}
\parbox{13cm}
{\caption[Delta]
{\sl The double scattering correction to gluon radiation in 
the rest frame of the target nucleus {\rm\bf (a)}. 
The absorptive part of the corresponding elastic $pA$ amplitude
{\rm\bf (b)}. The triple-Pomeron graph representing fusion $2\Pom \to 
\Pom$ {\rm\bf (c)}.}
\label{fusion}}
\end{center}
 \end{figure}

Although the two interpretations look so different, one can get a hint
that they are the same phenomenon relating them to the Reggeon graphs
which are Lorentz invariant. The double-scattering correction to the
cross section of gluon radiation depicted in Fig.~\ref{fusion}a
corresponds to the absorptive part of elastic $pA$ amplitude shown in
Fig.~\ref{fusion}b.
 Since the initial and final nucleons are colorless, each pair of
exchanged gluons attached to the same nucleon is in a colorless state,
i.e. represents the Pomeron (in the Born approximation). Thus, the
Feynman graph in Fig.~\ref{fusion}b is a part of the triple-Pomeron
diagram shown in Fig.~\ref{fusion}c. It can be interpreted as fusion of
two Pomerons originated from different nucleons, $2\Pom \to \Pom$. This
observation bridges the two interpretations of gluon shadowing.

Note that in the hadronic representation such a suppression of parton
density corresponds to Gribov's inelastic shadowing \cite{gribov} which 
is related to the single diffraction cross section. In
particular, gluon shadowing corresponds to the triple-Pomeron term in the
diffractive dissociation cross section which enters the calculations of
inelastic corrections (Fig.~\ref{fusion}c).

There are still very few numerical evaluations of gluon shadowing in the
literature, all done in the rest frame of the nucleus using the idea from
\cite{al}. It turns out that even at low $Q^2$ in the nonperturbative
domain where one should expect the strongest shadowing effects they are
rather weak. Indeed, data for diffractive excitation of the incident
hadrons to the states of large mass, the so called triple-Pomeron region,
show that the cross section is amazingly small, an order of magnitude
smaller than one could expect by comparing with the cross section of
small mass excitation \cite{kklp}. To explain such a smallness one has to
assume a rather small radius of propagation of the LC gluons, $r_0\approx
0.3\,\fm$ \cite{kst2,k3p}. It is clear that such a small quark-gluon
fluctuation also leads to a rather weak gluon shadowing.

To incorporate the smallness of the size of quark-gluon fluctuations into
the LC dipole approach a nonperturbative LC potential describing the
quark-gluon interaction was introduced into the Schr\"odinger equation
for the LC Green function describing the propagation of a quark-gluon
system.  The strength of the potential was fixed by data on high mass
($M_X^2$) diffraction $pp\to pX$ \cite{kst2}. This approach allows to
extend the methods of pQCD to the region of small $Q^2$. Since a new
semihard scale $1/r_0 \sim 0.65\,\GeV$ is introduced one should not
expect a substantial variation of gluon shadowing at $Q^2 \lsim 4/r_0^2$.
Indeed, calculations performed in \cite{kst2} for $Q^2=0$ and $4\,\GeV^2$
using different techniques led to about the same gluon shadowing. At
higher $Q^2$ shadowing slowly (logarithmically) decreases in accordance
with expectations based on the evolution equation \cite{mq}.

We repeated the calculations \cite{kst2} of the ratio of the gluon
densities in nuclei and nucleon,
 \beq
R_G(x_{Bj},Q^2)=\frac{G_A(x_{Bj},Q^2)}{A\,G_N(x_{Bj},Q^2)} \approx
1 - \frac{\Delta\sigma(\bar qqG)}
{\sigma_{tot}^{\gamma^*A}}\ ,
\label{RG}
 \eeq
 where $\Delta\sigma(\bar qqG)$ is the inelastic correction to the total
cross section $\sigma_{tot}^{\gamma^*A}$ related to the creation of a
$\bar qqG$ intermediate state,
 \beqn 
&& \Delta\sigma(\bar qqG) = 
{\rm Re}\int\limits_{-\infty}^{\infty} 
dz_2 \int\limits_{-\infty}^{z_2} dz_1\,
\rho_A(b,z_1)\,\rho_A(b,z_2)
\int d^2x_2\,d^2y_2\,d^2x_1\,d^2y_1 \int
d\alpha_q\,\frac{d\,\alpha_G}{\alpha_G}
\nonumber\\ &\times&
F^{\dagger}_{\gamma^*\to\bar qqG}
(\vec x_2,\vec y_2,\alpha_q,\alpha_G)\
G_{\bar qqG}(\vec x_2,\vec y_2,z_2;\vec x_1,\vec y_1,z_1)\
F_{\gamma^*\to\bar qqG}
(\vec x_1,\vec y_1,\alpha_q,\alpha_G)\ .
\label{delta-sigma}
 \eeqn
 Here $\vec x$ and $\vec y$ are the transverse distances from the gluon
to the quark and antiquark, respectively. $\alpha_q$ is the fraction of
the LC momentum of the $\bar qq$ carried by the quark, and $\alpha_G$ is
the fraction of the photon momentum carried by the gluon.
$F_{\gamma^*\to\bar qqG}$ is the amplitude of diffractive $\bar qqG$
production in a $\gamma^*N$ interaction \cite{kst2},
 \beqn 
F_{\gamma^*\to\bar qqG}(\vec x,\vec y,\alpha_q,\alpha_G)
&=& {9\over8}\,
\Psi_{\bar qq}(\alpha_q,\vec x -\vec y)\,
\left[\Psi_{qG}\left(\frac{\alpha_G}{\alpha_q},
\vec x\right) - \Psi_{\bar qG}
\left(\frac{\alpha_G}{1-\alpha_q},\vec y\right)\,
\right]\nonumber\\ &\times& 
\biggl[\sigma_{\bar qq}(x)+
\sigma_{\bar qq}(y) - 
\sigma_{\bar qq}(\vec x - \vec y)\biggr]\ , 
\label{amplitude}
\eeqn
 where $\Psi_{\bar qq}$ and $\Psi_{\bar qG}$ are the LC distribution 
functions of the $\bar qq$ fluctuations of a photon and $qG$ fluctuations 
of a quark, respectively.

$G_{\bar qqG}(\vec x_2,\vec y_2,z_2;\vec x_1,\vec y_1,z_1)$ is the LC
Green function which describes propagation of the $\bar qqG$ system from
the initial state with longitudinal and transverse coordinates $z_1$ and
$\vec x_1,\vec y_1$, respectively, to the final coordinates $(z_2,\vec
x_2,\vec y_2)$. For the calculation of gluon shadowing one should
suppress the intrinsic $\bar qq$ separation, i.e. to assume $\vec x =
\vec y$. In this case the Green function essentially simplifies and
describes propagation of a gluon-gluon dipole through a medium.

An important finding of Ref.~\cite{kst2} is the presence of a strong
nonperturbative interaction which squeezes the gluon-gluon wave packet
and substantially diminishes gluon shadowing. The smallness of the
gluon-gluon transverse separation is not a model assumption, but is
dictated by data for hadronic diffraction to large masses (triple-Pomeron
regime) which is controlled by diffractive gluon radiation.

\begin{figure}[tbh]
\includegraphics{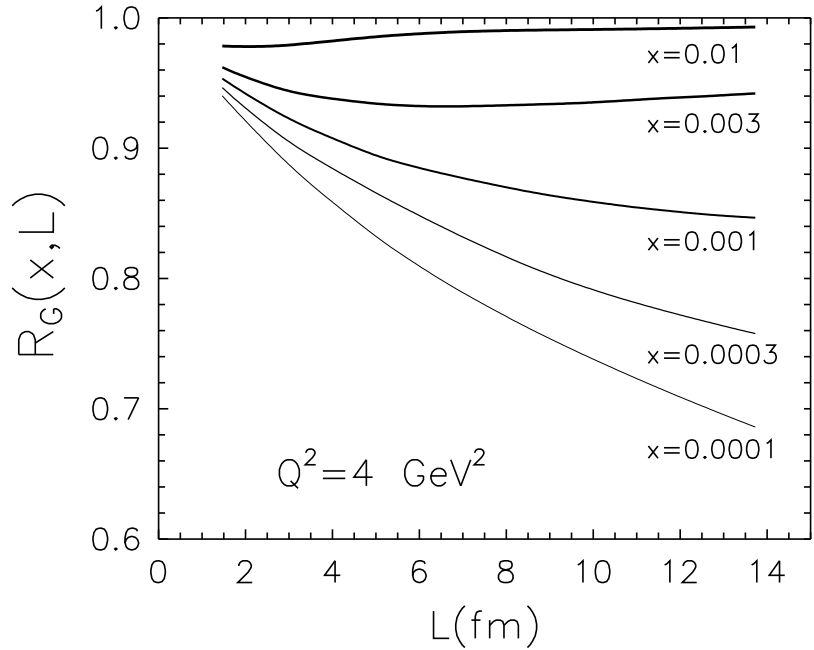}
\begin{center}
\vspace{9cm}
\parbox{13cm}
{\caption[Delta]
{\sl 
The ratio of nucleus to nucleon gluon densities as function of the 
thickness of the nucleus, $L=T(b)/\rho_0$, at $Q^2=4$ and different fixed 
values of $x_{Bj}$.} 
\label{glue-shad}}
\end{center}
 \end{figure}

Further calculational details can be found in \cite{kst2}. Here we
calculate $R_G$ [Eq.~(\ref{RG})] for different nuclear thicknesses
$T_A(b)$. Since we use an approximation of constant nuclear density (see
\ref{a}), $T_A(b)=\rho_0\,L$, where $L=2\,\sqrt{R_A^2-b^2}$, the ratio
$R_G(x_{Bj},Q^2)$ is also implicitly a function of $L$. An example for
the calculated $L$-dependence of $R_G(x_{Bj},Q^2)$ at $Q^2=4\,\GeV^2$ is
depicted in Fig.~\ref{glue-shad} for different values of $x_{Bj}$.
As one should expect, the longer $L$, the stronger is gluon shadowing 
at small $x_{Bj}$.

We calculated the gluon shadowing only for the lowest Fock component
containing just one LC gluon. In terms of the parton model it reproduces
the effects of fusion of many gluons to one gluon (in terms of Regge
approach it corresponds to $n\Pom \to \Pom$ vertex).  Inclusion of higher
multigluon Fock components is still a challenge. However, their effect
can be essentially taken into account by eikonalization of the calculated
$R_G(x_{Bj},Q^2)$, as argued in \cite{kth}. In other words, the dipole
cross section which is proportional to the gluon density at small
separations, should be renormalized everywhere,
 \beq
\sigma_{\bar qq} \Rightarrow 
R_G\,\sigma_{\bar qq}\ .
\label{1100}
 \eeq Such a procedure makes the nuclear medium more transparent. This
could be expected since Gribov's inelastic shadowing is known to suppress
the total hadron-nucleus cross sections, i.e. to make nuclei more
transparent \cite{zkl,kn}.

It is interesting, that the cross section of incoherent electroproduction
of vector mesons is rather insensitive to gluon shadowing. Indeed,
although the renormalization Eq.~(\ref{1100}) suppresses the
pre-exponential factor $\sigma_{\bar qq}(r,s)$ on the r.h.s. of
Eq.~(\ref{510}), it simultaneously increases the exponential. These two
effects essentially cancel. Indeed, our predictions for the effect of
gluon shadowing for incoherent production $\gamma^* A\to \rho^0 X$
depicted in Fig.~\ref{glue-incoh} demonstrate a rather small difference
between the curves with (solid) and without (dashed) gluon shadowing.

 A few observations are in order. First, the onset of gluon shadowing
happens at rather high energy $\nu > 100\,\GeV$. This corresponds to the
claim made in \cite{kst2} that the onset of gluon shadowing requires
smaller $x_{Bj}$ than the onset of quark shadowing. This is because the
fluctuations containing gluons are in general heavier than the $\bar qq$
and have a shorter CL.

Then, one can see that a stronger effect of gluon shadowing is expected
for nitrogen than for lead. Although it contradicts simple intuition, it
is easy interpreted. The renormalization of the dipole cross section
Eq.~(\ref{1100}) may either suppress or enlarge the incoherent cross
section in Eq.~(\ref{510}) depending on the value of the nuclear
thickness function $T(b)$. Namely, it should lead to a suppression for
small $T(b)$, but to an enhancement for large $T(b)$. Indeed we observe
this trend in Fig.~\ref{glue-incoh}. Some enhancement (antishadowing) can
be seen for lead at $Q^2=0$. The results presented for nitrogen show that
the effect is maximal at intermediate values of $Q^2$ while it is smaller
at $Q^2=0$ and $10\,\GeV$. This is a result of the same interplay between
the pre-exponential factor and the exponential in Eq.~(\ref{510}).

The implication of gluon shadowing for the case of coherent production
$\gamma^*A \to VA$ is clearer. It is easy to understand that it always
diminishes the coherent cross section. In terms of VDM the
photoproduction cross section is related to the total $VA$ cross section
which is always reduced by inelastic corrections. One can also see from
Eq.~(\ref{615}) that the suppression of the dipole cross section by the
renormalization [Eq.~(\ref{1100})] can result only in a reduction of the
cross section.  Thus, we expect much stronger effects of gluon shadowing
for coherent than for incoherent production. The results of calculations
based on the exact expression Eq.~(\ref{610}) confirm this conjecture.
The predicted suppression of the coherent cross section is stronger for
heavy nuclei and low $Q^2$.

\section{Summary and conclusions}\label{conclusions}

Electroproduction of vector mesons off nuclei is subject to an interplay
between coherence (shadowing) and formation (color transparency) effects.
Conventionally, one can associate those effects with the initial and
final state interaction respectively. We developed a rigorous
quantum-mechanical approach based on the light-cone QCD Green function
formalism which naturally incorporates these interference effects.  Our
main results and observations are the following.

\begin{itemize}

\item The suggested approach allows to find for the first time a
comprehensive model for the long standing problem of exclusive
electroproduction of vector mesons off nuclei. The main result of the
paper, Eq.~(\ref{520}), interpolates between the previously known low and
high energy limits for incoherent production. Eq.~(\ref{560})  does the
same for coherent production.

\item The onset of coherence effects (shadowing) can mimic the expected
signal of CT in incoherent electroproduction of vector mesons at medium
and low energies. In order to single out the formation effect data must
be binned in $l_c$ and $Q^2$. Observation of a rising nuclear
transparency as function of $Q^2$ for fixed $l_c$ would signal color
transparency.

\item Due to quark-hadron duality the Green function formalism under
consideration is equivalent to a solution of the full multi-channel
problem in the hadronic representation. We found a much steeper $Q^2$
dependence of nuclear transparency (Fig.~\ref{lc-const-inc}), i.e. a
stronger signal of CT, than was predicted in \cite{hk-97} within the
two-channel approximation.  Moreover, the slope of the $Q^2$ dependence
is even higher at lower energies. This should allow to detect a signal of
CT in experiments with the HERMES spectrometer and especially at JLab.

\item The successful experimental confirmation \cite{lc-hermes} of the
predicted coherence length effects \cite{hkn} seems now to be a
accidental consequence of the specific correlation between $Q^2$ and
$l_c$ in the HERMES data. The present parameter-free calculations well
describe the observed variation of nuclear transparency with $l_c$
(Fig.~\ref{lc-hermes-fig}) as a result of a complicated interplay between
the effects of CT and coherence length.

\item There are other effects which may cause a rise of nuclear
transparency with $Q^2$ at $l_c=const$ thus mimicking a signal of CT.  
These are the lowest order inelastic corrections which are rather
precisely fixed by available data for diffraction, and the finite
lifetime of vector mesons. Both effects lead to the more transparent
nuclear medium at higher energies, i.e. at larger $Q^2$ due to the
correlation in Eq.~(\ref{534}) between $\nu$ and $Q^2$. We found,
however, both effects to be too weak (Figs.~\ref{in-corr} and
\ref{decay-q2}) to be relevant.

\item The effects of CT in coherent production of vector mesons are found
to be less pronounced. Although transparency decreases with $Q^2$ and
does not mimic CT in this case, the CL effects significantly modify the
$Q^2$ dependence and may completely eliminate any signal of CT at medium
energies.  Besides, the cross section of coherent production is very
small at low energies (Fig.~\ref{glue-coh}).

\item The effects of CT modify the impact-parameter dependence of the
amplitude of coherent production by diminishing the mean square of the
impact parameter in the interaction amplitude. Therefore the positions of
the diffractive minima in the differential cross section are expected to
shift to larger values of $|t|$ (Fig.~\ref{t-dep-coh}). However the
effects seems to be too small to be reliably observed.

\item Although it has been known how to calculate nuclear transparency in
the high energy limit $l_c\gg R_A$ \cite{kz-91,knnz}, the effect of gluon
shadowing was missed. We calculated nuclear suppression of gluons at
small $x_{Bj}$ within the same LC approach treating it as shadowing
corrections for the higher Fock states containing gluons. The
nonperturbative interaction of the LC gluons significantly reduces the
predicted magnitude of gluon shadowing (Fig.~\ref{glue-shad}). Although
the amplitude of meson production off a bound nucleon is suppressed due
to a reduced amount of gluons in the nucleus, the same effect makes the
nuclear medium more transparent and enhances the meson survival
probability. For incoherent $\rho$ production these two effects nearly
compensate each other for heavy nuclei (Fig.~\ref{glue-incoh}).  The
cross section for coherent production is less for more transparent
nuclei, therefore the effect of gluon shadowing is more pronounced
(Fig.~\ref{glue-coh}). These corrections are not important at HERMES or
JLab energies, but are significant at the higher energies of eRHIC and
for coherent Coulomb production in heavy ion collisions at RHIC.

\end{itemize}

Concluding, the predicted large effects of CT in incoherent
electroproduction of vector mesons off nuclei open new possibilities for
the search for CT with medium energy electrons.

\medskip

\noindent
 {\bf Acknowledgment}:  We are grateful to Alexander Borissov and Kawtar
Hafidi for inspiring questions. J.N. is thankful to B.~Povh for
hospitality at Max-Planck-Institut f\"ur Kernphysik where this paper was
finished. This work has been partially supported by a grant from the
Gesellschaft f\"ur Schwerionenforschung Darmstadt (GSI), grant
No.~GSI-OR-SCH, and by the European Network:  Hadronic Physics with
Electromagnetic Probes, Contract No.~FMRX-CT96-0008. The work of J.N. has
been supported in part by the Slovak Funding Agency, Grant No. 2/1169 and
Grant No. 6114.

 \def\appendix{\par
 \setcounter{section}{0} \setcounter{subsection}{0}
 \def\thesection{Appendix \Alph{section}}
\def\thesubsection{\Alph{section}.\arabic{subsection}}
\def\theequation{\Alph{section}.\arabic{equation}}
\setcounter{equation}{0}}

 \appendix

\section{Approximation for the dipole cross section}\label{a}

To keep the calculations simple we are forced to use the approximate
dipole cross section Eq.~(\ref{460}) which allows to obtain the Green
function in an analytical form as is described in Sect.~\ref{green-f}. We
fix the factor $C(s)$ by demanding that calculations employing the
approximation Eq.~(\ref{460}) reproduce correctly the results based on
the realistic cross section in the limit $l_c\gg R_A$ when the Green
function takes the simple form (\ref{465}). Thus, for incoherent
production of vector mesons the factor $C(s)$ is fixed by the relation,
 \beqn
&& \frac{
\int d^{2}{b}\,T_A(b)\,\left|
\int d^{2}r\,r^2\,\exp\biggl[ - {1\over2}\, 
C_{T,L}(s)\,r^2\,T_{A}({b})\,\biggr]
\int d\alpha\,
\Psi_{V}^{*}(\vec r,\alpha)\,
\Psi^{T,L}_{\gamma^{*}}(\vec r,\alpha)
\right|^2 }
{\left|\int d^{2}r\,r^2\int d\alpha\,
\Psi_{V}^{*}(\vec r,\alpha)\,
\Psi^{T,L}_{\gamma^{*}}(\vec r,\alpha)\right|^2 }  
\nonumber\\ &=&
\frac{
\int d^{2}{b}\,T_A(b)\,\left|
\int d^{2}r\,\sigma_{\bar qq}(r,s)\,
\exp\biggl[ - {1\over2}\, 
\sigma_{\bar qq}(r,s)\,T_{A}({b})\,\biggr]
\int d\alpha\,
\Psi_{V}^{*}(\vec r,\alpha)\,
\Psi^{T,L}_{\gamma^{*}}(\vec r,\alpha)
\right|^2 }
{\left|\int d^{2}r\,\sigma_{\bar qq}(r,s)\int d\alpha\,
\Psi_{V}^{*}(\vec r,\alpha)\,
\Psi^{T,L}_{\gamma^{*}}(\vec r,\alpha)\right|^2 } 
\label{A20}
 \eeqn

Correspondingly, for coherent production the factor $C(s)$ is fixed by 
the relation,
 \beqn
&& \frac{
\int d^{2}b\,\left|\int d^{2}r\,\int d\alpha\,
\Psi_{V}^{*}(\vec r,\alpha)\,
\Psi^{T,L}_{\gamma^{*}}(\vec r,\alpha)\,
\left\{1 - \exp\biggl[ - {1\over2}\, 
C_{T,L}(s)\,r^2\,T_{A}({b})\,\biggr]
\right\}\right|^2 }
{\left|\int d^{2}r\,\int d\alpha\,
\Psi_{V}^{*}(\vec r,\alpha)\,C_{T,L}(s)\,r^2\,
\Psi^{T,L}_{\gamma^{*}}(\vec r,\alpha)\right|^2}  
\nonumber\\ &=&
\frac{
\int d^{2}{b}\,\left|\int d^{2}r\,\int d\alpha\,
\Psi_{V}^{*}(\vec r,\alpha)\,
\Psi^{T,L}_{\gamma^{*}}(\vec r,\alpha)\,
\left\{1 - \exp\biggl[ - {1\over2}\, 
\sigma_{\bar qq}(r,s)\,T_{A}({b})\biggr]
\right\}\right|^2}{\left|\int d^{2}r\,\int d\alpha\,
\Psi_{V}^{*}(\vec r,\alpha)\,\sigma_{\bar qq}(r,s)\,
\Psi^{T,L}_{\gamma^{*}}(\vec r,\alpha)\right|^2 }\, ,
\label{A30}
 \eeqn

To take advantage of the analytical form of the Green function which is
known only for the LC potential Eq.~(\ref{440}) with a constant nuclear
density, we use the approximation $\rho_{A}({b},z)  =
\rho_{0}\,\Theta(R_{A}^{2} - {b}^{2} - z^{2})$. Therefore we have to use
this form for Eqs.~(\ref{A20}) and (\ref{A30}) as well. The value of the
mean nuclear density $\rho_{0}$ has been determined using the relation,
 \BE
\int d^{2}{b}\,\biggl [ 1 - exp \biggl ( - \sigma_{0}\,
\rho_{0}\,\sqrt{R_{A}^{2} - {b}^{2}}\biggr ) \biggr ] =
\int d^{2}{b}\,\biggl [ 1 - exp 
\biggl ( - \frac{\sigma_{0}}{2}\,T({b})
\biggr ) \biggr ]\ ,
\label{A40}
 \EE
 where the nuclear thickness function $T_A(b)$ is calculated with the
realistic Wood-Saxon form of the nuclear density. The value of $\rho_{0}$
turns out to be practically independent of the cross section $\sigma_{0}$
in the range from 1 to 50 mb.


\begin{thebibliography}{99}

\bibitem{gribov} V.N.~Gribov, {\sl Sov. Phys. JETP} {\bf 29}, 483 (1969)
[Zh. Eksp. Teor. Fiz. {\bf 56}, 892 (1969)].

\bibitem{zkl} A.B.~Zamolodchikov, B.Z.~Kopeliovich and L.I.~Lapidus, {\sl
Pis'ma Zh. Eksp. Teor. Fiz.} {\bf 33}, 612 (1981);  {\sl Sov. Phys. JETP
Lett.} {\bf 33}, 595 (1981).

\bibitem{bbgg} G.~Bertsch, S.J.~Brodsky, A.S.~Goldhaber and J.F.~Gunion,
{\sl Phys. Rev. Lett.} {\bf 47}, 297 (1981).

\bibitem{jpr} P.~Jain, B.~Pire and J.P.~Ralston, Phys. Rept. {\bf 
271}, 67 (1996). 

\bibitem{gs} J.F.~Gunion and D.E.~Soper, {\sl Phys. Rev.} {\bf D15}, 2617
(1977).

\bibitem{hp} J.~H\"ufner and B.~Povh, {\sl Phys. Rev.} {\bf D46}, 990
(1992).

\bibitem{p} B.~Povh, {\sl Hadron Interactions - Hadron Sizes}, {\bf
hep-ph/9806379}

\bibitem{kz-91} B.Z.~Kopeliovich and B.G.~Zakharov, {\sl Phys. Rev.} {\bf
D44}, 3466 (1991).

\bibitem{hk-97} J.~H\"ufner and B.Z.~Kopeliovich, {\sl Phys. Lett.} {\bf
B403}, 128 (1997).

\bibitem{bauer} T.H.~Bauer et al., {\sl Rev. Mod. Phys.} {\bf 50}, 261
(1978).

\bibitem{nmc} The NMC Collaboration, M.~Arneodo et al., {\sl Nucl. Phys.}
{\bf 441}, 12 (1995);  {\sl Nucl. Phys.} {\bf 481}, 3 (1996).

\bibitem{e665} The E665 Collaboration, M.R.~Adams et al., {\sl Z. Phys.}
{\bf C67}, 403 (1995).

\bibitem{k-neutrino} B.Z.~Kopeliovich, {\sl Phys. Lett.} {\bf B227}, 461
(1989).

\bibitem{e772} The E772 Collaboration, D.M.~Alde et al., {\sl Phys. Rev.
Lett.} {\bf 64}, 2470 (1990).
 
\bibitem{e866} The E866 Collaboration, M.~Vasiliev et al., {\sl Phys.
Rev. Lett.} {\bf 83}, 2304 (1999).

\bibitem{hkn} J.~H\"ufner, B.Z.~Kopeliovich and J.~Nemchik, {\sl Phys.
Lett.} {\bf B383}, 362 (1996).

\bibitem{kk} V.~Karmanov and L.A.~Kondratyuk, {\sl Sov. Phys. JETP Lett.}
{\bf 18}, 266 (1973).

\bibitem{kklp} Yu.M.~Kazarinov, B.Z.~Kopeliovich, L.I.~Lapidus and
I.K.~Potashnikova, {\sl JETP} {\bf 70}, 1152 (1976).

\bibitem{al} A.H.~Mueller, {\sl Nucl. Phys.} {\bf B335}, 115 (1990).

\bibitem{krt2} B.Z. Kopeliovich, J.~Raufeisen and A.V.~Tarasov, {\sl
Phys. Rev.} {\bf C62}, 035204 (2000).

\bibitem{levin} E.~Gotsman, E.~Levin and U.~Maor, Phys. Lett. {\bf 425}, 
369 (1998)

\bibitem{gbw} K.~Golec-Biernat and M.~W\" usthoff, {\sl Phys. Rev.} {\bf
D59}, 014017 (1999); {\sl Phys. Rev.} {\bf D60}, 114023 (1999).

\bibitem{kst2} B.Z.~Kopeliovich, A.~Sch\" afer and A.V.~Tarasov, {\sl
Phys. Rev.} {\bf D62}, 054022 (2000).

\bibitem{dl} A.~Donnachie and P.V.~Landshoff, {\sl Phys. Lett.} {\bf
B478}, 146 (2000).

\bibitem{pdt} Review of Particle Physics, Particle Data Group, {\sl Eur.
Phys. J.} {\bf C15}, 1 (2000).

\bibitem{gribov-nonplanar} V.N.~Gribov,  Sov. J. Nucl. Phys. 
{\bf 9}, 369 (1969) [Yad. Fiz. {\bf 9}, 640 (1969)].

\bibitem{bronzan} J.B.~Bronzan, G.L.~Kane and U.P.~Sukhatme, {\sl
Phys.Lett.} {\bf B49} (1974) 272.

\bibitem{lc} J.B.~Kogut and D.E.~Soper, {\sl Phys. Rev.} {\bf D 1}, 2901
(1970).

\bibitem{bks-71} J.M.~Bjorken, J.B.~Kogut and D.E.~Soper, {\sl Phys.
Rev.} {\bf D 3}, 1382 (1971).

\bibitem{nz-91} N.N.~Nikolaev and B.G.~Zakharov, {\sl Z. Phys.} {\bf
C49}, 607 (1991).

\bibitem{fg} R.P.~Feynman and A.R.~Gibbs, {\sl Quantum Mechanics and Path
Integrals}, McGraw-Hill Book Company, NY 1965.

\bibitem{gamma1} H1 Collaboration, S.~Aid et al., {\sl Z. Phys.} {\bf
C69}, 27 (1995).

\bibitem{gamma2} ZEUS Collaboration, M.~Derrick et al., {\sl Phys. Lett.}
{\bf B293}, 465 (1992).

\bibitem{terentev} M.V.~Terent'ev, {\sl Sov. J. Nucl. Phys.} {\bf 24},
106 (1976).

\bibitem{hz} I.~Halperin and A.~Zhitnitsky, {\sl Phys. Rev.} {\bf D56}, 184
(1997).

\bibitem{jan97} J.~Nemchik, N.N.~Nikolaev, E.~Predazzi and B.G.~Zakharov,
{\sl Z. Phys.} {\bf C75}, 71 (1997).

\bibitem{nmc-phirho-q2}
NMC Collaboration, M.~Arneodo et al.,
{ Nucl. Phys.} {\bf B429}, 503 (1994).

\bibitem{H1-rho} H1 Collaboration, C.~Adloff et al., {\sl Eur. Phys. J.}
{\bf C13}, 371 (2000).

\bibitem{ZEUS-rho} ZEUS Collaboration, J.~Breitweg et al., {\sl Eur.
Phys. J.} {\bf C6}, 603 (1999).

\bibitem{q2-slope-phi-hera}
H1 Collaboration, C.~Adloff et al.,
{ Z. Phys.} {\bf C75}, 607 (1997); \\ 
H1 Collaboration, C.~Adloff et al.,
{ Phys. Lett} {\bf B483}, 360 (2000); \\ 
ZEUS Collaboration, M.~Derrick et al.,
{ Phys. Lett.} {\bf B380}, 220 (1996); \\
ZEUS Collaboration, J.~Breitweg et al.,
``Exclusive Electroproduction of $\Phi$ Mesons at HERA'',
presented at the XXIX International Conference on
HEP, Vancouver, 1998, paper-793.

\bibitem{jan98} J.~Nemchik, N.N.~Nikolaev, E.~Predazzi, B.G.~Zakharov and
V.R.~Zoller, {\sl J. Exp. Theor. Phys.} {\bf 86}, 1054 (1998).

\bibitem{q2-slope-rho-lowenergy}
CHIO Collaboration, W.D.~Shambroom et al.,
{Phys. Rev.} {\bf D26}, 1 (1982); \\
E665 Collaboration, M.R.~Adams et al.,
{Z. Phys.} {\bf C74}, 237 (1997).

\bibitem{H1-slope-a} 
H1 Collaboration, S.~Aid et al., 
{\sl Nucl. Phys.} {\bf B463}, 3 (1996). 

\bibitem{H1-slope-b} 
H1 Collaboration, S.~Aid et al., 
{\sl Nucl. Phys.} {\bf B468}, 3 (1996).

\bibitem{ZEUS-slope-a}
ZEUS Collaboration, M.~Derrick et al., 
{\sl Z. Phys.} {\bf C69}, 39 (1995).

\bibitem{ZEUS-slope-b}
ZEUS Collaboration, J.~Breitweg et al., 
{\sl Z. Phys.} {\bf C73}, 253 (1997); \\
ZEUS Collaboration, J.~Breitweg et al., 
{\sl Eur. Phys. J.} {\bf C2}, 247 (1998).

\bibitem{ZEUS-slope-c}
ZEUS Collaboration, M.~Derrick et al., 
{\sl Phys. Lett.} {\bf B356}, 601 (1995).

\bibitem{real-phi-slope}
J.~Busenitz et al.,{\sl Phys. Rev.} {\bf D40}, 1 (1989), and
references therein; \\
R.~Erbe et al., {\sl Phys. Rev.} {\bf 175}, 1669 (1968); \\ 
J.~Ballam et al., {\sl Phys. Rev.} {\bf D7}, 3150 (1973); \\
H.J.~Bersh et al., {\sl Nucl. Phys.} {\bf B70}, 257 (1974); \\
H.J.~Behrend et al., {\sl Phys. Lett.} {\bf B56}, 408 (1975);\\ 
D.P.~Barber et al., {\sl ibid.} {\bf B79}, 150 (1978); \\
D.~Aston et al., {\sl Nucl. Phys.} {\bf B172}, 1 (1980); \\
M.~Atkinson et al., {\sl Z. Phys.} {\bf C27}, 233 (1985).

\bibitem{ZEUS-phi-0}
ZEUS Collaboration, 
M.~Derrick et al., {\sl Phys. Lett.} {\bf B377}, 259 (1996).

\bibitem{le-rho-0a}
OMEGA Collaboration, D.~Aston et al., 
{\sl Nucl. Phys.} {\bf B209}, 56 (1982).

\bibitem{le-rho-0b}
J.~Park et al.,
{\sl Nucl. Phys.} {\bf B36}, 404 (1972); \\
R.M.~Egloff et al., 
{\sl Phys. Rev. Lett.} {\bf 43}, 657 (1979).

\bibitem{ZEUS-rho-0} 
ZEUS Collaboration, M.~Derrick et al., {\sl Z.
Phys.} {\bf C63}, 391 (1994); \\
M.~Derrick et al., {\sl paper} {\bf pa02-050},
submitted to the 28th International Conference on HEP, 25-31 July, 1996,
Warsaw, Poland.

\bibitem{real-rho-lowenergy}
W.G.~Jones et al., {\sl Phys. Rev. Lett.} {\bf 21}, 586 (1968); \\
C.~Berger et al., {\sl Phys. Lett.} {\bf B39}, 659 (1972); \\ 
SBT Collaboration, J.~Ballam et al., {\sl Phys. Rev.} {\bf D5}, 545 (1972); \\
G.E.~Gladding et al., {\sl ibid.} {\bf 8}, 3721 (1973).







\bibitem{krt1} B.Z.~Kopeliovich, J.~Raufeisen and A.V.~Tarasov, {\sl
Phys. Lett.} {\bf B440}, 151 (1998).

\bibitem{wolfram} T.~Renk, G.~Piller and W.~Weise, Nucl. Phys. {\bf
A689}, 869 (2001). 

\bibitem{hkz} J.~H\"ufner, B.Z.~Kopeliovich and A.~Zamolodchikov, {\sl Z.
Phys.} {\bf A357}, 113 (1997).

\bibitem{knnz} B.Z.~Kopeliovich, J.~Nemchik, N.N.~Nikolaev and
B.G.~Zakharov, {\sl Phys. Lett.} {\bf B324}, 469 (1994).

\bibitem{mueller} A.H.~Mueller, in Proc. of the 17th Recontre de Moriond,
Moriond. 1982, ed. by J.~Tran~Thanh Van, p. 13
 
\bibitem{brodsky} S.J.~Brodsky, {\sl in Proc. of the 13th Symposium on
Multiparticle Dynamics}, ed. by W.~Kittel, W.~Metzger and A.~Stergiou
(World Scientific, Singapore, 1982)

\bibitem{jk} B.K.~Jennings and B.Z.~Kopeliovich, Phys. Rev. Lett. {\bf
70}, 3384 (1993).

\bibitem{e665-rho} The E665 Collaboration, M.R.~Adams et al., {\sl Phys.
Rev. Lett.} {\bf 74}, 1525 (1995).

\bibitem{ktv}  B.Z.~Kopeliovich, A.V.~Tarasov and O.O.~Voskresenskaya,
{\sl Long-Range Coulomb Forces in DIS: Missed Radiative Corrections?},
hep-ph/0105110, to appear in Eur. Phys. J. {\bf A}.

\bibitem{kn95} B.Z.~Kopeliovich and J.~Nemchik, {\sl Where is the
Baseline for Color Transparency Studies with Moderate Energy Electron
Beams ?}, preprint {\bf MPIH-V41-1995};  {\bf nucl-th/9511018}.

\bibitem{lc-hermes} HERMES Collaboration, K.~Ackerstaff et al., {\sl
Phys. Rev. Lett.} {\bf 82}, 3025 (1999).

\bibitem{borissov} A.~Borissov, {\sl private communication}

\bibitem{jm} B.K. Jennings, G.A.~Miller, Phys. Rev. {\bf D44}, 692 
(1991).

\bibitem{murthy} P.V.R.~Murthy et al., Nucl. Phys. {\bf B92}, 269 (1975).

\bibitem{gsponer} A.~Gsponer et al., Phys. Rev. Lett. {\bf 42}, 9 (1979).

\bibitem{kn} B.Z.~Kopeliovich and J.~Nemchik, {\sl Phys. Lett.} {\bf
B368}, 187 (1996).

\bibitem{glr} L.V.~Gribov, E.M.~Levin and M.G.~Ryskin, Nucl. Phys. {\bf 
B188} (1981) 555; Phys. Rep. {\bf 100}, 1 (1983).

\bibitem{k3p} B.Z.~Kopeliovich, I.K.~Potashnikova, B.~Povh and
E.~Predazzi, {\sl Phys. Rev. Lett.} {\bf 85}, 507 (2000).

\bibitem{mq} A.H.~Mueller and J.W.~Qiu, Nucl. Phys. {\bf B268}, 427 
(1986).

\bibitem{kth} B.Z.~Kopeliovich, A.V.~Tarasov and J.~H\"ufner,
{\sl Coherence Phenomena in Charmonium Production off Nuclei at the 
Energies of RHIC and LHC}, hep-ph/0104256, to appear in Nucl. Phys. {\bf 
A}.

\end{thebibliography}
\end{document}